\documentclass[preprint,12pt]{elsarticle}
\usepackage[T1]{fontenc}
\usepackage{amsmath,amssymb,amsthm,mathtools,bm}
\usepackage{graphicx,booktabs,array,multirow}
\usepackage{adjustbox}
\usepackage{flafter}

\usepackage{algorithm,algpseudocode}
\usepackage{xcolor}
\usepackage[colorlinks=true,linkcolor=blue!50!black,citecolor=blue!50!black,urlcolor=blue!50!black]{hyperref}
\biboptions{numbers,sort&compress}
\journal{Computer Methods in Applied Mechanics and Engineering}
\makeatletter
\def\ps@pprintTitle{%
  \let\@oddhead\@empty\let\@evenhead\@empty
  \let\@oddfoot\@empty
  \let\@evenfoot\@oddfoot}
\makeatother

\newcommand{\R}{\mathbb{R}}
\newcommand{\X}{\mathcal{X}}
\newcommand{\A}{\mathcal{A}}
\newcommand{\U}{\mathcal{U}}
\newcommand{\Ob}{\mathcal{O}}
\newcommand{\E}{\mathcal{E}}
\newcommand{\B}{\mathcal{B}}
\newcommand{\Rr}{\mathcal{R}}
\newcommand{\psir}{\psi_{\mathrm{ref}}}
\newcommand{\Fr}{F_{\mathrm{ref}}}
\newcommand{\nP}[1]{\|#1\|_{P}}
\newcommand{\ipP}[2]{\langle #1,#2\rangle_{P}}
\newcommand{\SY}{S_{Y}}
\newcommand{\PSY}{P_{S_Y}}
\newcommand{\rF}{r_{F}}
\newcommand{\rhoF}{\rho_{F}}
\newcommand{\T}{^{\mathsf T}}
\newcommand{\e}[1]{\times10^{#1}}
\newcommand{\dd}{\,\mathrm{d}}
\DeclareMathOperator{\spn}{span}
\DeclareMathOperator{\diag}{diag}

\newtheorem{theorem}{Theorem}
\newtheorem{proposition}{Proposition}
\newtheorem{lemma}{Lemma}
\theoremstyle{definition}
\newtheorem{remark}{Remark}

\hypersetup{pdftitle={Constitutive-fibre mechanics: mechanics-directed uncertainty in constitutive function space},pdfauthor={Altaf Ahmad Lone, R. T. Durai Prabhakaran, Timon Rabczuk, Xiaoying Zhuang}}

\begin{document}
\widowpenalty=10000
\clubpenalty=10000
\displaywidowpenalty=10000

\begin{frontmatter}
\title{Constitutive-fibre mechanics: mechanics-directed uncertainty in constitutive function space}
\author[iitj]{Altaf Ahmad Lone}
\author[iitj]{R. T. Durai Prabhakaran}
\author[fud,buw]{Timon Rabczuk\corref{cor}}
\ead{timon.rabczuk@uni-weimar.de}
\author[luh]{Xiaoying Zhuang}
\cortext[cor]{Corresponding author}
\affiliation[iitj]{organization={Department of Mechanical Engineering, Indian Institute of Technology Jammu}, city={Jammu}, postcode={181221}, country={India}}
\affiliation[fud]{organization={Fudan University}, city={Shanghai}, country={China}}
\affiliation[buw]{organization={Institute of Structural Mechanics, Bauhaus-Universit\"at Weimar}, city={Weimar}, postcode={99423}, country={Germany}}
\affiliation[luh]{organization={Institute of Photonics, Department of Mathematics and Physics, Leibniz Universit\"at Hannover}, city={Hannover}, country={Germany}}
\begin{abstract}
A structural prediction rests on constitutive functions that are known only through a finite number of tests. The tests fix the response along the states they visit; together with a declared regularity bound they leave a set of admissible constitutive functions, which we call the constitutive fibre. Constitutive-fibre mechanics (CFM) computes the range of a quantity of interest over this set. Adjoint solves of the mechanical problem represent the test observations and the quantity of interest as elements of the constitutive function space. The part of the sensitivity of the quantity of interest that is orthogonal to all observation sensitivities is hidden from the tests; for exact data it alone determines the half-width of the prediction, and for data with tolerance the half-width is a finite convex problem in the span of the adjoint representers. The comparison of the two values shows whether a new type of test or a more precise test is needed, and a rank-one update ranks candidate tests before they are performed. A learned constitutive class restricts the fibre and can therefore only under-represent the range; the deficit is computable. Beyond the linear range, an optimality system with state-dependent representers gives the extreme constitutive functions, and under nominal data the half-width is odd and the midpoint even in the amplitude. The same set decides whether the prediction is defined for every constitutive function that the data admit, which is a different question from its precision. For a phase-field bar with unknown degradation function, $3\%$ of the sensitivity of the stability eigenvalue is hidden and produces $70\%$ of its uncertainty, and one reaction measured beyond the tested range reduces the half-width to $13\%$. For an open-hole composite lamina with coupon data, the coupon that sharpens the compliance most is not the one that makes the prediction valid, and validity at a fixed load is proved, by a certificate accepted in exact arithmetic, on a declared cell of the data-consistent set of a finite-dimensional model.
\end{abstract}
\end{frontmatter}

\section{Introduction}\label{sec:intro}

Every structural analysis uses constitutive functions: a strain-energy density, a degradation function, a yield surface. None of them is measured as a function. A test programme records a finite number of forces, strains or displacements, each with a tolerance, on specimens that visit a limited set of constitutive states. Between and beyond those states the constitutive function is fixed by the chosen model form and not by the data. A prediction for a structure that visits other states, at a hole edge, near a crack tip or close to a stability limit, therefore depends on constitutive information that the tests did not supply. This paper asks how large that dependence is for a given quantity of interest, which tests would reduce it, and whether the prediction is defined at all for every constitutive function that the tests admit.

\paragraph{Calibration, discovery and their uncertainty} The classical route selects a parametric model and identifies its parameters, from homogeneous tests or from full-field data with finite-element model updating and the virtual fields method \cite{avril2008}. Bayesian calibration turns measurement noise into a posterior on the parameters \cite{kennedy2001,stuart2010}, and sampling-based sensitivity analysis with surrogate models propagates parameter uncertainty to structural responses, including fracture properties computed with phase-field models \cite{vubac2016,hamdia2015}. In both cases the uncertainty is conditional on the model form. Recent work removes part of this restriction. Model discovery selects a sparse law from a large library \cite{flaschel2021}, with a hierarchical Bayesian version that carries uncertainty over the library coefficients \cite{joshi2022}. Constitutive neural networks build physical constraints such as objectivity and polyconvexity into the architecture \cite{linka2023,klein2022}, and Bayesian variants attach credible intervals to the learned response \cite{linka2025,anton2025}. Neural networks are also used to solve the variational problem of the mechanics itself, with the energy as loss function \cite{samaniego2020}, including phase-field fracture \cite{goswami2020}; in CFM they enter as constitutive representations, and Section~\ref{sec:repr} states what their spread can and cannot show. Constitutive relations are also learned from indirect observations by optimisation constrained by the balance laws, with adjoint sensitivities \cite{akerson2025}. In all these approaches the uncertainty lives on the parameters of one representation. A direction of the constitutive function that the representation cannot express carries no uncertainty, whether or not the data determine it.

\paragraph{Identifiability and experimental design} Whether indirect observations determine a constitutive function is the question of identifiability. Ju \cite{ju2026} shows for scalar state-dependent constitutive functions of several classes of field equations that sparse indirect observations of one type leave the posterior dominated by the prior however many are collected, and that a few direct measurements at well-placed states are decisive. Bayesian optimal experimental design chooses tests by expected information gain \cite{chaloner1995,huan2013,alexanderian2021}; it has been applied to the load path of a cruciform specimen \cite{ricciardi2023} and to history-dependent constitutive laws \cite{bhattacharya2026}. These criteria measure information about parameters. A structural prediction usually depends on a few specific combinations of the constitutive unknowns, and information that does not reach those combinations does not improve it.

\paragraph{Set-based descriptions} A different line of work describes what is known by a set. Model-free data-driven mechanics replaces the constitutive law by the material data set and solves for the state closest to it \cite{kirchdoerfer2016,conti2018}; it does not extrapolate beyond the data. Convex and interval models propagate bounded uncertainties through a structural analysis \cite{benhaim1990,moens2005}. Optimal uncertainty quantification computes extreme values of a quantity of interest over all models consistent with stated assumptions and information \cite{owhadi2013}. In approximation theory, optimal recovery and information-based complexity determine the smallest worst-case error with which a functional can be estimated from finitely many, possibly inaccurate, linear measurements of an element of a known class \cite{golomb1959,micchelli1977,melkman1979,traub1980,traub1988,plaskota1996,foucart2024}. Our companion work on constitutive-set mechanics (CSM) \cite{lone2026csm} follows the set-based view for the potential energy: the constitutive function ranges over an admissible set, and the extreme values of the minimum potential energy over that set are bounded and certified with a small number of finite-element solves.

\paragraph{The question of this paper} CSM answers how far the energy can move over a given admissible set. It does not say why the set is as large as it is, and it is tied to the energy as the quantity of interest. The present paper starts from the tests. Given the observations of a test programme with their tolerances, a reference law and a declared bound on the regularity of the deviation from it, the set of admissible constitutive functions that reproduce the observations is fixed. We call it the \emph{constitutive fibre}, since it is the preimage of the data under the observation map, and we call the resulting method constitutive-fibre mechanics (CFM). The object of the method is the range of a quantity of interest over the fibre. The quantity of interest is arbitrary: an energy, a reaction, a local strain, a stiffness or a stability eigenvalue.

\paragraph{Contributions} The paper makes the following contributions.
\begin{enumerate}
\item \emph{A geometry generated by the mechanics.} Adjoint solves represent each observation and each quantity of interest in the constitutive function space with the declared metric. The span of the observation representers contains the directions the tests see, and its orthogonal complement is hidden. For exact data the half-width of the prediction is the prior radius times the norm of the hidden part of the representer of the quantity of interest (Theorem~\ref{thm:exact}), which makes identifiability a quantitative property of a specific prediction.
\item \emph{Finite tolerance and the choice of test.} For data with tolerance the half-width is a finite convex problem in the span of the representers (Theorem~\ref{thm:tol}). Its excess over the exact-data value is what more precise tests of the existing types can remove; the exact-data value can only be reduced by a test that sees a new constitutive direction. A rank-one update ranks candidate tests in advance.
\item \emph{Finite amplitude.} The extreme laws satisfy an optimality system in which the representers move with the law (Theorem~\ref{thm:kkt}), and under nominal data the half-width is odd and the midpoint even in the amplitude (Theorem~\ref{thm:parity}).
\item \emph{Representations.} A parametrised or learned constitutive class restricts the fibre; it can under-represent the range and never certify it, and the deficit is computable.
\item \emph{Validity.} Whether the prediction is defined for every law of the fibre is a property of the same set as its precision, governed by different constitutive directions and improved by different tests. For a composite plate we prove it, on a cell of the data-consistent set of a finite-dimensional model, by a certificate accepted in exact arithmetic.
\end{enumerate}
The linear layer coincides with optimal recovery from inaccurate data (Section~\ref{sec:reading}); what CFM adds is the mechanical origin of the geometry, the assessment of representations, the finite-amplitude layer and the coupling of precision with validity.

\paragraph{Outline} Section~\ref{sec:fibre} defines the fibre and the prediction range. Section~\ref{sec:linear} develops the linear theory. Section~\ref{sec:nonlinear} treats finite amplitude, Section~\ref{sec:repr} representations, Section~\ref{sec:acq} the choice of the next experiment and Section~\ref{sec:validity} validity. Section~\ref{sec:alg} collects the algorithms. Section~\ref{sec:examples} presents three examples: a two-element bar that can be followed by hand, a phase-field bar with unknown degradation function, and an open-hole composite lamina with coupon data. Section~\ref{sec:discussion} discusses scope and limits. Proofs are given in \ref{app:proofs}, and the exact certification used for the plate in \ref{app:cert}.
\section{The constitutive fibre}\label{sec:fibre}

\subsection{Constitutive class, metric and admissibility}\label{sec:class}

Let $\psi$ denote the constitutive function of the problem at hand: a strain-energy density $\psi(e)$ on a strain domain, a degradation function $g(d)$ on $[0,1]$, or a collection of such functions. We write $\Omega_c$ for its domain, the set of constitutive states, to distinguish it from the physical domain of the structure. A reference law $\psir$ is given, for instance the calibrated model in current use, and the unknown law is written $\psi=\psir+v$ with $v$ in a Hilbert space $\X$ of functions on $\Omega_c$ with inner product $\ipP{\cdot}{\cdot}$ and norm $\nP{\cdot}$. Three ingredients describe what is assumed about $\psi$, and we keep them separate.
\begin{itemize}
\item \emph{Physics.} A closed convex set $\A$ collects the conditions that every law must satisfy: symmetries, a natural state, monotonicity or convexity where the material requires it, end conditions such as $g(0)=1$, $g(1)=0$. Conditions that are linear in $\psi$, including pointwise inequalities on $\psi$ and its derivatives, always give a convex $\A$.
\item \emph{Regularity.} The deviation from the reference is bounded in the declared metric, $\nP{\psi-\psir}\le K$. The metric is a Sobolev norm of sufficient order: if the mechanics evaluates derivatives of $\psi$ up to order $k$ at points of a $d$-dimensional state domain, point evaluation of these derivatives must be continuous on $\X$, which requires an order $s>k+d/2$.
\item \emph{Evidence.} Whatever is derived from measurements enters through the observations of Section~\ref{sec:obs}, with tolerances, and not through $\A$.
\end{itemize}
The metric and the radius $K$ are declared quantities. They state how fast the constitutive function is allowed to vary between tested states, and every result below is conditional on them; Section~\ref{sec:declare} gives a procedure for choosing them.

The set of admissible laws is a subset of the function space $\X$, and it is not parametrised. In computations $\X$ is replaced by a finite-dimensional subspace spanned by a numerical basis, such as splines or polynomials in the constitutive state. This basis is a discretisation in the sense of the finite element method: it approximates the function space, carries no assumption about the material, and is refined until the computed quantities no longer change. It must not be confused with a parametrised constitutive model, whose parameters define the admissible laws; the distinction is taken up in Section~\ref{sec:repr}.

\subsection{Mechanics, observations and quantities of interest}\label{sec:obs}

For a given law the mechanical problem is written abstractly as
\begin{equation}\label{eq:state}
\B(q;\psi)=0,
\end{equation}
where $q$ collects the state variables (displacements, phase field, load multipliers) of a finite-element model, and $\B$ is the residual of equilibrium or of an incremental stationarity condition. We consider a regular branch: $\partial_q\B$ is invertible, so that $q=q(\psi)$ locally. A test programme provides $m$ scalar observations
\begin{equation}
\Ob_i[\psi]=M_i\big(q_i(\psi),\psi\big),\qquad i=1,\dots,m,
\end{equation}
each computed on the model of its own specimen with state $q_i$; examples are reactions at prescribed displacements, gauge strains and stresses at measured strains. The measured values are $y\in\R^m$, and the measurement errors are assumed to lie in a closed convex set $\E\subset\R^m$ containing the origin, for instance a ball $\|\eta\|\le\varepsilon$ or a box of per-reading tolerances. The structural prediction is a vector of $p$ quantities of interest
\begin{equation}
F[\psi]=J\big(q_s(\psi),\psi\big)\in\R^p,
\end{equation}
evaluated on the model of the structure with state $q_s$.

\paragraph{Adjoint sensitivities} All derivatives with respect to the law are obtained by the adjoint method. For a scalar functional $J(q(\psi),\psi)$ with state equation \eqref{eq:state},
\begin{equation}\label{eq:adjoint}
DJ[\psi]\,v=\partial_\psi J\,v-\lambda\T\,\partial_\psi\B(q;\psi)\,v,\qquad (\partial_q\B)\T\lambda=(\partial_qJ)\T .
\end{equation}
One linear solve with the transposed tangent gives the derivative in every constitutive direction $v$. Since $\psi$ enters $\B$ through its values and derivatives at the constitutive states visited by the model, $DJ[\psi]$ is a distribution supported on those states: a sum of point evaluations of $v$ and its derivatives at the quadrature-point states, weighted by the adjoint. For a stability eigenvalue the formula involves one more derivative of $\psi$ than for an energy or a reaction, through the derivative of the tangent.

\subsection{Fibre and prediction range}

\begin{figure}[tp]
\centering
\includegraphics[width=\textwidth]{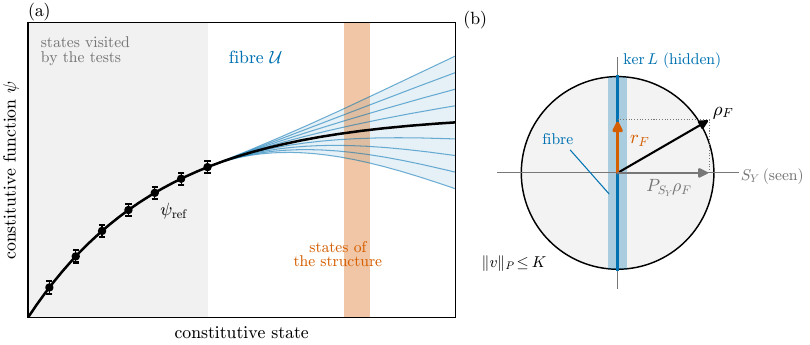}
\caption{The constitutive fibre and the split of the sensitivity of a quantity of interest. (a) The tests constrain the law only along the constitutive states they visit; the fibre $\U$ is the set of admissible laws within the regularity ball that reproduce the observations within tolerance, and the structure may visit states beyond the tested ones. (b) In the constitutive function space, the representer $\rhoF$ of the quantity of interest splits into a part in the span $\SY$ of the observation representers and a hidden part $\rF$. For exact data the fibre is the section of the ball by $\ker L=\SY^\perp$, and the half-width of the prediction is $K\nP{\rF}$.}
\label{fig:concept}
\end{figure}

The \emph{constitutive fibre} is the set of laws that the declared information admits,
\begin{equation}\label{eq:fibre}
\U=\Big\{\psi\in\A:\ \nP{\psi-\psir}\le K,\ \ \Ob[\psi]-y\in\E\Big\}.
\end{equation}
For a direction $a\in\R^p$ in the space of the quantities of interest, the support function of the image of the fibre,
\begin{equation}\label{eq:support}
H(a)=\sup_{\psi\in\U}a\T F[\psi],
\end{equation}
gives the largest value of $a\T F$ that the information allows. The prediction of $a\T F$ is the interval $[-H(-a),H(a)]$ with midpoint and half-width
\begin{equation}\label{eq:halfwidth}
c(a)=\tfrac12\big(H(a)-H(-a)\big),\qquad W(a)=\tfrac12\big(H(a)+H(-a)\big).
\end{equation}
For a single quantity of interest we write $F$, $\Fr=F[\psir]$ and $W$. Figure~\ref{fig:concept} illustrates the construction.

Three properties follow from the definition. First, the range is monotone: additional observations, tighter tolerances, a smaller radius or further admissibility conditions shrink $\U$ and therefore $W$, provided the error sets compose by intersection, as product sets and boxes do. Second, $W(a)$ is the smallest worst-case error with which $a\T F$ can be predicted from the declared information, and the midpoint attains it. Third, the range is a statement about the declared information and not about an unknown true law: if the true law satisfies the declarations, its value lies in the interval, and if it does not, the interval says nothing about it. We therefore call $W$ the uncertainty that the constitutive information permits and avoid the word error.

\subsection{Choosing the reference law, the metric and the radius}\label{sec:declare}

The three declarations have different roles, and each can be fixed by a procedure that does not use the quantity of interest.

\paragraph{Reference law} The reference $\psir$ is the centre of the regularity ball, not an estimate of the true law. The natural choices are the calibrated model in current use, a learned law, or the admissible law of smallest distance from a simple model that reproduces the data within tolerance. It should satisfy the physics, $\psir\in\A$, and preferably the data, $\Ob[\psir]-y\in\E$. A reference that misses the data is permitted, since the offset $b$ enters Theorem~\ref{thm:tol}, but the fibre then lies on one side of the ball and its size depends on how far the reference is from the data. The quantity $K_{\min}$ of Proposition~\ref{prop:gram} measures this distance.

\paragraph{Metric} The metric is chosen in three steps. First, its order follows from the mechanics: if derivatives of $\psi$ up to order $k$ enter the observations or the quantity of interest, the order $s$ must exceed $k+d/2$ (Section~\ref{sec:class}); we take the smallest integer order that satisfies this, or one more where boundary conditions of the function space make a higher order natural. Second, the constitutive states are scaled by the extent of $\Omega_c$, for instance by the failure strains, and the function by a reference value, so that the norm is dimensionless and all state directions are measured on the same footing. Third, the derivative orders are weighted equally unless there is physical information to the contrary. In practice the highest derivative dominates the norm of a smooth deviation, so the order and the scaling are the decisive choices.

\paragraph{Radius} The radius bounds the deviation of the law from the reference in the declared metric. Three considerations fix it. (i) It must not be smaller than $K_{\min}$, since a smaller radius contradicts the data. (ii) It should contain the laws that are considered plausible for the material class: published models, calibrations of other model forms to the same data, or a law fitted to additional curves. The largest distance of such laws from $\psir$ is the natural value of $K$; in Section~\ref{sec:plate} it is the distance of a nonlinear law fitted to published curves. (iii) The result is reported together with its dependence on $K$. For nominal data the half-width is homogeneous, $V(K,\varepsilon)=K\,V(1,\varepsilon/K)$, so it grows linearly with $K$ in the hidden-dominated regime and otherwise depends on the radius only through this factor and the ratio $\varepsilon/K$. Admissibility conditions also have a radius at which they become active, which Lemma~\ref{lem:activity} computes.

The quantities on which the decisions of CFM rest are less sensitive to these choices than the size of the range. The hidden fraction $\sin\theta$, the alignment $\cos(\rF,r_c)$ of a candidate test and the contraction \eqref{eq:cos} do not depend on $K$ at all; they depend on the mechanics and the metric only. Which test to perform next can therefore be decided before the radius is settled, and the radius then fixes the absolute size of the range.

\subsection{Relation to constitutive-set mechanics}

In CSM \cite{lone2026csm} the set of laws is given directly, the quantity of interest is the minimum potential energy $F[\psi]=\min_q\Pi(q;\psi)$, and its extreme values over the set are bounded by certified two-sided estimates. By Danskin's theorem \cite{danskin1967} the derivative of this $F$ needs no adjoint solve: $DF[\psi]v=\partial_\psi\Pi(q^\ast;\psi)v$ at the minimiser $q^\ast$, which is the occupation of the constitutive states used in CSM. CFM takes the set from the test programme through \eqref{eq:fibre}, admits any differentiable quantity of interest through \eqref{eq:adjoint}, and analyses the structure of the set in the directions that matter for that quantity.
\section{Linear theory: seen and hidden constitutive directions}\label{sec:linear}

\subsection{Representers}

Linearise the observations and the quantities of interest at the reference law, $L=D\Ob[\psir]:\X\to\R^m$ and $DF[\psir]:\X\to\R^p$. Each row is a bounded linear functional on $\X$ by the order condition of Section~\ref{sec:class}, so the Riesz theorem gives representers
\begin{equation}\label{eq:riesz}
L_iv=\ipP{\rho_i}{v},\qquad DF_j[\psir]\,v=\ipP{\rho_{F_j}}{v}\qquad\text{for all }v\in\X .
\end{equation}
Each representer costs one adjoint solve \eqref{eq:adjoint} followed by the solution of $R_P\rho=\ell$, where $R_P$ is the Riesz operator of the metric; in a discretisation of $\X$ this is one solve with the Gram matrix of the basis. For a direction $a$ we write $\rho_a=\sum_ja_j\rho_{F_j}$, and $\rhoF$ for a single quantity of interest. Let
\begin{equation}
\SY=\spn\{\rho_1,\dots,\rho_m\},\qquad \SY^\perp=\ker L .
\end{equation}
A constitutive perturbation in $\SY$ changes at least one observation to first order: the tests \emph{see} it. A perturbation in $\ker L$ changes none: it is \emph{hidden}. The representer of the quantity of interest splits accordingly,
\begin{equation}\label{eq:split}
\rhoF=\PSY\rhoF+\rF,\qquad \rF=(I-\PSY)\rhoF\in\ker L,
\end{equation}
with $\PSY$ the orthogonal projection onto $\SY$. We call $\rF$ the hidden part of the sensitivity and
\begin{equation}
\sin\theta=\nP{\rF}/\nP{\rhoF}
\end{equation}
the hidden fraction. The split depends on the mechanics of the tests, on the mechanics of the structure and on the metric; the subspace $\ker L$ itself depends on the mechanics of the tests alone.

In this section the fibre is replaced by its linearisation at $\psir$,
\begin{equation}\label{eq:linfibre}
C=\big\{v\in\X:\ \nP{v}\le K,\ \ Lv-b\in\E\big\},\qquad b=y-\Ob[\psir],
\end{equation}
and $F$ by $\Fr+DF[\psir]v$; admissibility is added in Section~\ref{sec:adm}. The data are called nominal if $b=0$.

\subsection{Exact data}

\begin{theorem}[exact data]\label{thm:exact}
Let $\E=\{0\}$ and $b=0$. Then the linearised range of $a\T F$ is symmetric about $a\T\Fr$ with half-width
\begin{equation}\label{eq:V0}
V_0(a)=K\,\nP{r_a},\qquad r_a=(I-\PSY)\rho_a .
\end{equation}
The image of the linearised fibre in the space of the quantities of interest is the ellipsoid $\Fr+K\,G_{\mathrm{hid}}^{1/2}\bar B$, where $\bar B$ is the closed unit ball of $\R^p$ and
\begin{equation}\label{eq:Ghid}
(G_{\mathrm{hid}})_{jk}=\ipP{r_{F_j}}{r_{F_k}}
\end{equation}
is the Gram matrix of the hidden parts.
\end{theorem}

The theorem gives identifiability a quantitative and target-specific meaning. The tests need not determine the constitutive function; it suffices that the representer of the quantity of interest lies in the span of the observation representers. Conversely, a law may be tightly constrained in most directions while a prediction remains uncertain because its sensitivity has a component along one of the unconstrained ones. For exact data with $b\ne0$ the fibre is the section of the ball by an affine subspace, and the same formula holds with $K^2$ replaced by $K^2-\nP{v_b}^2$ and the midpoint shifted by $\ipP{\rho_a}{v_b}$, where $v_b$ is the minimum-norm solution of $Lv=b$; this is the hypercircle of Golomb and Weinberger \cite{golomb1959}.

\begin{remark}[linearisation at exact data]\label{rem:rank}
The exact-data fibre of the nonlinear problem is tangent to $\ker L$ only where $D\Ob$ has constant rank. If the rank drops at $\psir$, the linearisation can be wrong at first order: for the single observation $\Ob[\psi]=x_1^2-x_2$ of a law with two parameters and $y=0$, the fibre is a parabola, while $\ker L$ at the origin is a line. With a positive tolerance this difficulty does not arise, and the finite-amplitude theory of Section~\ref{sec:nonlinear} does not use the tangent space.
\end{remark}

\subsection{Data with tolerance}

\begin{theorem}[finite tolerance]\label{thm:tol}
Let $\E$ be closed and convex with $-b$ in its interior, so that $v=0$ is strictly feasible in \eqref{eq:linfibre}. Then
\begin{equation}\label{eq:dual}
V_\E(a):=\sup_{v\in C}\ipP{\rho_a}{v}=\min_{\mu\in\R^m}\Big[K\,\nP{\rho_a-L^\ast\mu}+h_\E(\mu)+\mu\T b\Big],
\end{equation}
where $L^\ast\mu=\sum_i\mu_i\rho_i$ and $h_\E(\mu)=\sup_{\eta\in\E}\mu\T\eta$ is the support function of the error set. For the Euclidean ball of radius $\varepsilon$ and nominal data,
\begin{equation}\label{eq:Veps}
V_\varepsilon(a)=\min_{\mu}\Big[K\,\nP{\rho_a-L^\ast\mu}+\varepsilon\|\mu\|\Big]
=\inf_{\tau\ge0}\Big[(K^2+\tau\varepsilon^2)\,a\T G_\tau a\Big]^{1/2},
\end{equation}
with
\begin{equation}
(G_\tau)_{jk}=\ipP{\rho_{F_j}}{(I+\tau L^\ast L)^{-1}\rho_{F_k}}.
\end{equation}
A maximiser lies in $\spn\{\rho_a,\rho_1,\dots,\rho_m\}$, so the problem is finite-dimensional whatever the dimension of $\X$.
\end{theorem}

The minimisation over $\mu$ has a direct interpretation. The term $L^\ast\mu$ is the part of $\rho_a$ that is explained by the observations; it is paid for at the price of the tolerance, $h_\E(\mu)$, while the unexplained remainder is paid for at the price of the prior radius. As $\varepsilon\to0$ the multiplier is free and the remainder becomes $r_a$, so $V_\varepsilon\downarrow V_0$; as $\varepsilon\to\infty$, $\mu=0$ and $V_\varepsilon\to K\nP{\rho_a}$, the prior half-width. The one-parameter form in \eqref{eq:Veps} shows the intermediate structure: for each $\tau$ the fibre is contained in the ellipsoid $\nP{v}^2+\tau\|Lv\|^2\le K^2+\tau\varepsilon^2$, and the half-width is the smallest bound that these ellipsoids give. Since the optimal $\tau$ depends on the direction $a$, the image of the fibre is in general not an ellipsoid; $V_\varepsilon(a)$ is a support function, and for several quantities of interest it must be evaluated direction by direction. In the example of Section~\ref{sec:pf} the ratio of $V_\varepsilon(a)$ to the value suggested by the exact-data ellipsoid varies between $1.1$ and $3.1$ with the direction.

For a box of per-reading tolerances $\varepsilon_i$ the support function is $h_\E(\mu)=\sum_i\varepsilon_i|\mu_i|$ and \eqref{eq:dual} is a second-order cone problem. Boxes have the advantage that adding a reading can only shrink the fibre. A joint Euclidean ball whose radius is kept fixed while readings are added does not represent a fixed measurement precision.

\subsection{Gram form for several data blocks}\label{sec:gram}

Let the observations come in blocks $j=1,\dots,J$ with Euclidean tolerances $\varepsilon_j$ and offsets $b_j$, and stack the linearised quantities of interest and observations as $\mathcal D=(DF[\psir];L)$. All information is contained in the Gram matrix of the representers,
\begin{equation}
\Gamma=\mathcal D\,R_P^{-1}\mathcal D^\ast=\begin{pmatrix}\Gamma_{FF}&\Gamma_{FD}\\ \Gamma_{DF}&\Gamma_{DD}\end{pmatrix},\qquad \Gamma_{rs}=\ipP{\rho_r}{\rho_s}.
\end{equation}
For multipliers $\tau=(\tau_1,\dots,\tau_J)\ge0$ let $\Lambda_\tau=\diag(\tau_j^{-1}I)$ and
\begin{equation}\label{eq:master}
\begin{aligned}
m_\tau&=\Gamma_{FD}(\Gamma_{DD}+\Lambda_\tau)^{-1}b,\qquad
\Sigma_\tau=\Gamma_{FF}-\Gamma_{FD}(\Gamma_{DD}+\Lambda_\tau)^{-1}\Gamma_{DF},\\
c_\tau&=b\T(\Gamma_{DD}+\Lambda_\tau)^{-1}b .
\end{aligned}
\end{equation}

\begin{proposition}[Gram form]\label{prop:gram}
The linearised fibre is non-empty if and only if $K\ge K_{\min}$ with $K_{\min}^2=\sup_{\tau\ge0}\big[c_\tau-\sum_j\tau_j\varepsilon_j^2\big]$. If it has an interior point, then
\begin{equation}\label{eq:Vgram}
\sup_{v\in C}a\T DF[\psir]v=\inf_{\tau\ge0}\Big\{a\T m_\tau+\Big[\big(K^2+\textstyle\sum_j\tau_j\varepsilon_j^2-c_\tau\big)\,a\T\Sigma_\tau a\Big]^{1/2}\Big\}.
\end{equation}
\end{proposition}

The quantity $K_{\min}$ is the norm of the smallest deviation from the reference that reproduces the data within tolerance. A declared radius below $K_{\min}$ contradicts the data, which is a useful consistency check on the prior. The expressions \eqref{eq:master} are numerically delicate when the observation Gram matrix $\Gamma_{DD}$ is ill-conditioned, as it is for closely spaced readings; we evaluate them through an orthogonal factorisation of the representers and never through the normal equations.

\subsection{Admissibility}\label{sec:adm}

Let the part of $\A$ that can become active near the reference be described by linear inequalities $\ipP{\gamma_t}{v}\le m_t$, $t\in T$, where $m_t>0$ is the margin of constraint $t$ at $\psir$; a pointwise convexity condition, for example, gives one such inequality for every state and direction. The same duality argument gives
\begin{equation}\label{eq:admdual}
V_{\E,\A}(a)=\min_{\mu,\ \nu\ge0}\Big[K\,\nP{\rho_a-L^\ast\mu-\textstyle\sum_t\nu_t\gamma_t}+h_\E(\mu)+\mu\T b+\textstyle\sum_t\nu_tm_t\Big],
\end{equation}
with $\nu$ finitely supported. Admissibility can only reduce the half-width, and with it the range ceases to be symmetric even for nominal data. Whether a constraint matters is decided by the same calculus.

\begin{lemma}[activity]\label{lem:activity}
On the linearised fibre without constraint $t$, the smallest margin of that constraint is $m_t-V_\E(\gamma_t)$. The constraint is redundant if $V_\E(\gamma_t)<m_t$, and it can be active only if $V_\E(\gamma_t)\ge m_t$.
\end{lemma}

A physical condition is thus tested for relevance by treating its own functional as a quantity of interest. In the phase-field example of Section~\ref{sec:pf}, monotonicity and positivity of the degradation function are redundant, with $\max_tV_\E(\gamma_t)/m_t=0.18$. In the composite example of Section~\ref{sec:plate}, convexity of the strain energy is active and halves the half-width of the compliance.

\subsection{Reading the result}\label{sec:reading}

\paragraph{Two numbers and a decision} For one quantity of interest the linear analysis returns the prediction $\Fr$ and two half-widths, $V_0\le V_\varepsilon$. The first is caused by constitutive directions that no test of the programme sees, however precisely it is measured. The difference is caused by the tolerance of the tests that do see the relevant directions. If $V_\varepsilon\approx V_0$, repeating or refining the existing tests cannot help, and a test of a new type is needed, one whose representer has a component along $\rF$. If $V_\varepsilon\gg V_0$, the existing types of test suffice and their precision or number should be increased. Section~\ref{sec:acq} turns this reading into a ranking of candidate tests.

\paragraph{Relation to optimal recovery} With $\A=\X$ the linearised problem is the recovery of a linear functional of an element of a Hilbert ball from inaccurate linear information, the central problem of optimal recovery \cite{micchelli1977,traub1980}. The exact-data fibre is the hypercircle \cite{golomb1959}; $V_\E(a)$ is the radius of information of the functional; the one-parameter family in \eqref{eq:Veps} is the family of regularised estimators that Melkman and Micchelli proved optimal \cite{melkman1979}, see also \cite{plaskota1996,foucart2024}; and for a Sobolev metric the representers are splines whose knots are the constitutive states visited by the mechanics \cite{kimeldorf1971,wahba1990}. We claim none of this as new. What the mechanical setting contributes is the construction of the information and target functionals by adjoint solves of the test and structure models, the admissibility calculus, and the layers of the following sections, which have no counterpart in the linear theory: representations, finite amplitude and validity.

\paragraph{Relation to Gaussian-process conditioning} If $v$ is given the Gaussian prior whose covariance operator is $R_P^{-1}$, then $\Gamma$ is the prior covariance of the linearised outputs, and for a fixed $\tau$ the quantities $m_\tau$ and $\Sigma_\tau$ in \eqref{eq:master} are the posterior mean and covariance of $F$ given the data with noise covariance $\Lambda_\tau$. A linear Bayesian analysis with a chosen noise level evaluates one member of the family. The CFM half-width is the envelope \eqref{eq:Vgram} over the admissible noise levels, scaled by the set radius; it is a bound over a set, and no probability is attached to it. The two descriptions share the hidden directions: where $\rF\ne0$, the posterior variance of $F$ does not vanish as the noise tends to zero.
\section{Finite amplitude}\label{sec:nonlinear}

The linear theory replaces the observation map and the quantity of interest by their first-order expansions at the reference. For a radius $K$ and tolerance of practical size the laws at the ends of the range differ visibly from the reference, and the states of the tests and of the structure move with them. This section treats the support problem \eqref{eq:support} without linearisation, for one direction $a$ and with the Euclidean error ball; we write $f=a\T F$.

\subsection{Optimality system}

An end point of the range solves
\begin{equation}\label{eq:nlp}
\max_{\psi\in\A}\ f[\psi]\quad\text{subject to}\quad \tfrac12\nP{\psi-\psir}^2\le\tfrac12K^2,\qquad \tfrac12\|\Ob[\psi]-y\|^2\le\tfrac12\varepsilon^2 .
\end{equation}
At every law $\psi$ the adjoint method gives representers $\rho_f[\psi]$ and $\rho_i[\psi]$ of $Df[\psi]$ and $D\Ob_i[\psi]$ as in \eqref{eq:riesz}, computed on the states that this law produces.

\begin{theorem}[optimality with moving span]\label{thm:kkt}
Let $\psi^\ast$ be a local maximiser of \eqref{eq:nlp} in the interior of $\A$ at which both constraints are active and their gradients are linearly independent. Then there are multipliers $\alpha\ge0$, $\beta\ge0$ such that
\begin{equation}\label{eq:kkt}
\rho_f[\psi^\ast]-\sum_{i=1}^m\mu_i\,\rho_i[\psi^\ast]-\alpha\,(\psi^\ast-\psir)=0,\qquad \mu=\beta\,\big(\Ob[\psi^\ast]-y\big).
\end{equation}
If admissibility constraints $\ell_t(\psi)\le0$ are active at $\psi^\ast$, the term $-\sum_t\nu_t\gamma_t$ with $\nu_t\ge0$ is added on the left.
\end{theorem}

Equation \eqref{eq:kkt} is the finite-amplitude form of the split \eqref{eq:split}: the deviation of the extreme law from the reference is, up to the factor $\alpha$, the part of the sensitivity of the quantity of interest that the observations do not explain, with all sensitivities evaluated at the extreme law itself. The span of the observation representers moves with $\psi^\ast$, and \eqref{eq:kkt} is a nonlinear equation in function space. For $\alpha>0$ it shows that $\psi^\ast-\psir$ lies in the span of $m+1$ representers, as in the linear case. We solve it by a sequential quadratic programme in which every iteration costs one state solve and one adjoint solve per observation and per quantity of interest (Algorithm~\ref{alg:nl}).

\paragraph{Second-order check} Let $\mathcal L=f-\tfrac12\beta\|\Ob-y\|^2-\tfrac12\alpha\nP{\psi-\psir}^2$ be the Lagrangian and $T$ the tangent space of the two active constraints at $\psi^\ast$. If
\begin{equation}\label{eq:sosc}
c:=\alpha-\lambda_{\max}\Big(P_T\,R_P^{-1}D^2\big(f-\tfrac12\beta\|\Ob-y\|^2\big)[\psi^\ast]\,P_T\Big)>0,
\end{equation}
then $\psi^\ast$ is a strict local maximiser, and $c$ is the curvature of the constrained problem in the metric of $\X$. The margin $c$ is computable from second-order adjoints or from second differences of state solves.

\paragraph{What the end points are} A solution of \eqref{eq:kkt} is a law of the fibre: it is admissible, it lies in the ball, and it reproduces the data within tolerance, all of which are checked directly. Its value of $f$ is therefore attained, and the interval between the computed lower and upper end points is contained in the true range. The reverse inclusion would require a global bound on $f$ over the fibre. We establish local optimality by \eqref{eq:sosc} and report the end points as attained values, not as certified global extremes; a rigorous outer enclosure at finite amplitude is discussed in Section~\ref{sec:discussion}.

\subsection{Parity of midpoint and half-width}

Scale the declared information by an amplitude $t>0$, $(K,\varepsilon)=t\,(K_0,\varepsilon_0)$, with nominal data $y=\Ob[\psir]$. Let $f^+(t)$ and $f^-(t)$ be the upper and lower end points and $\bar V$ the linear half-width at $(K_0,\varepsilon_0)$ from Theorem~\ref{thm:tol}.

\begin{theorem}[parity]\label{thm:parity}
Let $f$ and $\Ob$ be smooth near $\psir$, let $\psir$ lie in the interior of $\A$, and let the linear problem at $(K_0,\varepsilon_0)$ have a unique maximiser satisfying strict complementarity and the second-order sufficient condition. Then for small $t$
\begin{equation}\label{eq:parity}
\begin{aligned}
\tfrac12\big(f^+(t)+f^-(t)\big)-f[\psir]&=s_2t^2+s_4t^4+\dots,\\
\tfrac12\big(f^+(t)-f^-(t)\big)&=\bar V\,t+s_3t^3+\dots
\end{aligned}
\end{equation}
\end{theorem}

The midpoint has no linear term and the half-width has no quadratic term. The linear half-width $\bar Vt$ is therefore accurate to third order in the amplitude, while the centre of the range departs from the reference prediction at second order. In practical terms, the size of the uncertainty can be read from the linear theory far beyond the range in which the response is linear in the law, whereas the location of the interval requires the nonlinear end points once $s_2t^2$ is comparable with the half-width. The proof in \ref{app:proofs} uses only the symmetry of the scaled problem under $(t,v)\mapsto(-t,-v)$; active admissibility or an offset in the data break this symmetry and with it the parity.
\section{Representations and learned constitutive classes}\label{sec:repr}

Data-driven constitutive modelling represents the law in a finite class, such as a polynomial, a spline or a neural network, fits it to the data, and often reports the spread of its fitted members as uncertainty. Two uses of a finite basis must be distinguished. As a discretisation of $\X$ (Section~\ref{sec:class}), a basis is a computational approximation that is refined until the computed range no longer changes; the error it introduces is a numerical error, of the same kind as the error of a finite-element mesh. As a constitutive model, a class is a statement about the material, which excludes all laws it cannot express however fine the computation. The mathematics below covers both uses, since both restrict the fibre; what differs is that the first restriction is removed by refinement and the second is not. In the present setting a class has a precise role. Let $\A_\Theta\subset\A$ be the set of laws that the class can represent and $\U_\Theta=\U\cap\A_\Theta$ the part of the fibre inside it.

\begin{proposition}[inner approximation]\label{prop:inner}
For every direction $a$,
\begin{equation}
H^\Theta(a):=\sup_{\psi\in\U_\Theta}a\T F[\psi]\ \le\ H(a).
\end{equation}
If the class is a linear space containing $\psir$, with tangent space $T\subset\X$ and orthogonal projection $P_T$, then the Gram matrix of the projected representers satisfies $\Gamma^\Theta\preceq\Gamma$, and for exact nominal data
\begin{equation}\label{eq:deficit}
V_0(a)^2-V_0^\Theta(a)^2=K^2\,\nP{r_a-P_{T\cap\ker L}\,\rho_a}^2\ \ge0 .
\end{equation}
\end{proposition}

The statement is elementary, and its consequences are practical. A representation can only under-report the range of a prediction. Agreement of all fitted members of a class on a prediction shows that the class has no freedom in the relevant hidden direction; it does not show that the data determine the prediction. Through $\Gamma^\Theta\preceq\Gamma$ a restricted class has three effects: it reduces the half-width (under-coverage), it changes the midpoint when the data are not nominal (bias), and it can make the fibre empty within the class although the data are consistent with the prior (incompatibility). A class can therefore never certify a range, but it can be assessed: the coverage ratio $V^\Theta/V$ is computable from the representers and measures, for the prediction at hand, how much of the admissible range the class can express.

\paragraph{What a class must resolve} By \eqref{eq:deficit} the deficit vanishes when the class contains the hidden part $r_a$. Since $r_a$ is a combination of the representers of the quantity of interest and of the observations, and these are splines with knots at the constitutive states visited by the structure and by the tests, a good class resolves those states. A class that is uniform over the whole constitutive domain, or adapted to the tested states only, spends its freedom where the prediction has no sensitivity. Section~\ref{sec:pf} shows differences of four orders of magnitude between classes of equal dimension.

\paragraph{Nonlinear classes} For a nonlinearly parametrised class, such as a neural network, the support problem \eqref{eq:nlp} is solved over the parameters, and any feasible member gives a lower bound on $H$. Near a nondegenerate end point $\psi^\ast$ with margin $c$ from \eqref{eq:sosc}, a feasible law $\psi_\Theta$ with both constraints active satisfies
\begin{equation}\label{eq:valueloss}
f[\psi^\ast]-f[\psi_\Theta]\ \ge\ \tfrac12\,c\,\nP{e_T}^2+o\big(\nP{e}^2\big),\qquad e=\psi_\Theta-\psi^\ast,
\end{equation}
with $e_T$ the tangential part of $e$. The value of the quantity of interest is thus a second-order measure of the error of the law: a class can reproduce the end-point value closely while its extreme law is visibly different from the true extreme law. A fit score of the prediction should not be read as a fit of the constitutive function.
\section{Selecting the next experiment}\label{sec:acq}

\subsection{Rank-one update}

Let a candidate test provide one further scalar observation with representer $\rho_c$, computed by an adjoint solve on the model of the candidate specimen at the reference law. Only its hidden part $r_c=(I-\PSY)\rho_c$ carries new information. For exact data the new hidden part of the quantity of interest is $r_F-\ipP{r_F}{r_c}\,r_c/\nP{r_c}^2$, so that
\begin{equation}\label{eq:cos}
\frac{V_0^{\mathrm{new}}}{V_0}=\sqrt{1-\cos^2(\rF,r_c)},\qquad \cos(\rF,r_c)=\frac{\ipP{\rF}{r_c}}{\nP{\rF}\nP{r_c}} .
\end{equation}
For several quantities of interest the hidden Gram matrix \eqref{eq:Ghid} is updated by
\begin{equation}\label{eq:downdate}
G_{\mathrm{hid}}^{\mathrm{new}}=G_{\mathrm{hid}}-\frac{c\,c\T}{\nP{r_c}^2},\qquad c_j=\ipP{r_{F_j}}{r_c},
\end{equation}
which is the quotient property of Schur complements \cite{crabtree1969}. A candidate is useful in proportion to the alignment of its hidden part with the hidden part of the prediction. A test whose representer lies in $\SY$ is a repetition of existing information, whatever its nominal novelty, and a test with a large hidden part orthogonal to $\rF$ teaches much about the law and nothing about the prediction. For a tolerance $\varepsilon_c$ on the candidate, the post-test half-width is obtained by solving \eqref{eq:dual} with the row added; this costs one small convex problem per candidate.

\subsection{The outcome of the test is not needed}

The half-width after the test depends on its outcome $y_c$, which is unknown when the test is selected. Within the linear theory this dependence is benign.

\begin{proposition}[nominal outcome]\label{prop:outcome}
For the linearised fibre, the post-test half-width $W(y_c)$ is a concave function of the outcome. If the fibre before the test is symmetric about the reference, as it is for nominal data and inactive admissibility, then $W(y_c)\le W(0)$: the nominal outcome is the least favourable one.
\end{proposition}

The half-width computed at the nominal outcome is thus a guaranteed post-test bound, and candidates can be ranked before any of them is performed. This is the classical observation that adaptive information is no better than nonadaptive information for linear problems on symmetric convex sets \cite{traub1980,traub1988}. Adaption can pay only through what breaks the symmetry: nonlinearity of the mechanics, offsets in the data, or active admissibility constraints. Section~\ref{sec:pf} verifies the predicted contraction against the nonlinear end points.

\subsection{Limits of information}\label{sec:limits}

Three statements bound what any extension of a test programme can achieve. Consider a family of possible observations $\{\ell_\theta\}$ with representers $\rho_\theta$, let $S_\infty$ be the closed span of all of them and $H_\infty=S_\infty^\perp$.

\begin{proposition}[information, precision, coverage]\label{prop:limits}
\begin{enumerate}
\item[(i)] For nested designs whose representers become dense in $S_\infty$ and tolerances tending to zero, the half-width converges to $K\nP{P_{H_\infty}\rho_a}$. It tends to zero if and only if $\rho_a\in S_\infty$.
\item[(ii)] If every reading has a tolerance of at least $\varepsilon>0$ and $M=\sup_\theta\nP{\rho_\theta}<\infty$, then for every design $V(a)\ge\nP{\rho_a}\min(K,\varepsilon/M)$.
\item[(iii)] If all observation sensitivities are supported in a closed set $\mathcal V\subset\Omega_c$ of visited constitutive states, every $v\in\X$ that vanishes on $\mathcal V$ together with the derivatives that the mechanics evaluates is hidden. The half-width is then at least $K$ times the norm of the projection of $\rho_a$ on these functions, for any number and precision of readings.
\end{enumerate}
\end{proposition}

Statement (i) says that completeness of the information is needed only in the direction of the target. Statement (ii) says that the number of readings cannot compensate for finite precision. Statement (iii) is the one with the clearest mechanical content: a constitutive function is nonlinear in the state, and information about it is information about the states that the specimens visit. Complete boundary data on a specimen do not give complete constitutive information if the specimen does not reach the states that the structure reaches. The lower bound in (iii) is the \emph{coverage floor}; it is computed from the sensitivity of the quantity of interest alone and vanishes when the tests visit all states to which the prediction is sensitive.

Together with Section~\ref{sec:reading}, these statements give the decision rule of CFM. If the half-width is dominated by $V_0$, and in particular if $V_0$ is close to its coverage floor, the programme needs a test that visits new states or measures a new kind of response. If it is dominated by the tolerance, statement (ii) and the dual \eqref{eq:dual} identify the readings whose precision matters: those with the largest $\varepsilon_i|\mu_i|$ at the optimum.
\section{Validity of a prediction}\label{sec:validity}

\subsection{The prediction as a partially defined map}

So far the quantity of interest was assumed to be defined on the whole fibre. In general it is not. A constitutive function is declared on a bounded domain $\Omega_c$ of states, the only region where the tests, the metric and the admissibility conditions say anything about it. The structural problem \eqref{eq:state} for a law $\psi$ may have no equilibrium whose states lie in $\Omega_c$, or the equilibrium may lie on a singular branch. For a fixed structure and load we therefore define the regular domain
\begin{equation}
\Rr=\big\{\psi\in\A:\ \eqref{eq:state}\ \text{has a regular solution}\big\},
\end{equation}
where a solution $q$ is regular if all its constitutive states lie in the interior of $\Omega_c$ and $\partial_q\B(q;\psi)$ is invertible,
and regard the prediction $F$ as a map defined on $\Rr$ only. Two different questions can then be asked of a fibre $\U$:
\begin{itemize}
\item \emph{validity}: is $\U\subset\Rr$, so that the prediction is defined for every law that the information admits;
\item \emph{precision}: how large is the image $F(\U)$, which is the subject of Sections~\ref{sec:linear} to \ref{sec:acq} and is meaningful once validity holds.
\end{itemize}
Both are properties of the same set, and both improve when the fibre shrinks. They are nevertheless governed by different constitutive directions. Precision is limited by the hidden part of the sensitivity of the quantity of interest. Validity is limited by the laws of the fibre that drive some state of the structure towards the boundary of $\Omega_c$, typically the most compliant admissible response in the most loaded region.

\subsection{Set-level statements}

The following statements need no linearisation, convexity or parametrisation; their proofs are short and given in \ref{app:proofs}. Their value is that they organise mechanics, representation and test design around one object.

\begin{proposition}[restriction]\label{prop:Arestr}
If $\emptyset\ne\U'\subset\U$, then $\U\subset\Rr$ implies $\U'\subset\Rr$ and $F(\U')\subset F(\U)$. In particular, for a constitutive class $\A_\Theta$ the set of loads at which $\U\cap\A_\Theta$ is valid contains the set of loads at which $\U$ is valid.
\end{proposition}

A representation removes laws from the fibre. It can remove exactly those that obstruct validity, and so make the validity range of a prediction appear larger than it is, in the same way as it makes the uncertainty appear smaller (Proposition~\ref{prop:inner}). Restriction by information and restriction by a representation have the same geometry and different justification: correct information retains the true law under the declared error model, while a representation retains it only if that is separately argued.

\begin{proposition}[what a test can guarantee]\label{prop:Bgain}
Let a candidate test assign to every $\psi\in\U$ a non-empty set $Y_c(\psi)$ of possible outcomes, and let $\U_{c,y}=\{\psi\in\U:y\in Y_c(\psi)\}$ be the fibre after outcome $y$. Then all post-test fibres are valid if and only if $\U$ is valid.
\end{proposition}

No test can guarantee, for every outcome, that an invalid prediction becomes valid: a law of $\U\setminus\Rr$ is retained by the outcomes it produces. A favourable outcome can establish validity; what a design can control is which outcomes do. Precision behaves differently, since the diameter of $F(\U_{c,y})$ involves pairs of laws retained by the same outcome, and it can be reduced for all outcomes (Proposition~\ref{prop:outcome}).

\begin{proposition}[pair separation]\label{prop:Cpair}
Fix a precision $\delta>0$ and call a pair of laws in $\U$ consequential if exactly one of them lies in $\Rr$, or if both do and their values of $F$ differ by more than $\delta$. Every non-empty post-test fibre is either disjoint from $\Rr$, or contained in $\Rr$ with $\operatorname{diam}F(\U_{c,y})\le\delta$, if and only if $Y_c(\psi_1)\cap Y_c(\psi_2)=\emptyset$ for every consequential pair.
\end{proposition}

For additive errors in a ball of radius $\varepsilon_c$, the condition reads $\|\Ob_c[\psi_1]-\Ob_c[\psi_2]\|>2\varepsilon_c$. Proposition~\ref{prop:Cpair} places validity and precision in one framework: a test is characterised by the pairs of laws it separates, and the pairs that matter for validity are in general not those that matter for precision. Since continuous observations with non-zero tolerance cannot separate laws that are arbitrarily close on opposite sides of the boundary of $\Rr$, a practical design works with a declared margin. In the plate example of Section~\ref{sec:plate}, the coupon that reduces the range of the compliance most leaves the validity margin violated, and the coupons that restore it reduce the range less.

\begin{proposition}[complete information]\label{prop:Dcomplete}
Let $\U_n$ be nested, non-empty and compact in a topology in which $\Rr$ is open and $F$ is continuous on $\Rr$, and let $\U_\infty=\bigcap_n\U_n\subset\Rr$. Then $\U_N\subset\Rr$ for some finite $N$, and $F(\U_n)\to F(\U_\infty)$ in the Hausdorff distance.
\end{proposition}

If the limiting image is a single point the prediction becomes exact; uniqueness of the constitutive function is sufficient for this and not necessary. Propositions~\ref{prop:limits} and \ref{prop:Dcomplete} describe the complete-information limit, in which hidden directions disappear. CFM is concerned with the regime before that limit, which is the regime of every actual test programme.

\begin{proposition}[prior information alone]\label{prop:Eprior}
Let the structure be loaded by prescribed forces proportional to a load $P$, let $\A$ be a cone, as it is when only symmetries, a natural state and convexity are imposed, and let $\psir\in\A$ with $K\ge\nP{\psir}$. If $\Omega_c$ is bounded, the prior set $\{\psi\in\A:\nP{\psi-\psir}\le K\}$ is not contained in $\Rr$ for any $P\ne0$.
\end{proposition}

The prior set contains the laws $\alpha\psir$, $0<\alpha\le1$, and the structure with law $\alpha\psir$ under load $P$ has the states of the structure with law $\psir$ under load $P/\alpha$. Since the stresses of $\psir$ are bounded on a bounded domain, these states leave $\Omega_c$ for small $\alpha$. Convexity excludes negative stiffness and not small stiffness. Validity must therefore come from the data, through readings that bound the stiffness from below along the states that the structure loads, or from an additional declared bound. A regularity prior that is wide enough to be credible cannot provide it.

\subsection{Certifying validity}\label{sec:gate}

Establishing $\U\subset\Rr$ requires an existence statement for every law of a set, which sampling cannot give. We use an a posteriori existence argument of Newton--Kantorovich type \cite{kantorovich1982,nakao2019}, made uniform over cells of the fibre. Let a cell $\mathcal C\subset\U$ be given together with a predictor $p(\psi)$ of the state, for instance a low-order polynomial in the coordinates of the law. Let $A$ be a fixed symmetric positive definite operator on the state space with energy norm $\|\cdot\|_A$, and suppose that for all $\psi\in\mathcal C$ and all states $q$ in the ball $\bar B_A(p(\psi),R)$ the states lie in $\Omega_c$ and the tangent is bounded below, $\partial_q\B(q;\psi)\succeq\kappa A$ with $\kappa>0$.

\begin{lemma}[gate]\label{lem:gate}
If $\|\B(p(\psi);\psi)\|_{A^{-1}}<\kappa R$ for all $\psi\in\mathcal C$, then for every $\psi\in\mathcal C$ the problem \eqref{eq:state} has exactly one solution in $\bar B_A(p(\psi),R)$; hence $\mathcal C\subset\Rr$.
\end{lemma}

The inequality compares the equilibrium residual of the predictor with the \emph{capacity} $\kappa R$ of the ball. It has to hold on a continuum of laws. When the law space is discretised by polynomials, the residual of a polynomial predictor is a polynomial in these coordinates, and the gate becomes the non-negativity of a polynomial on a semi-algebraic set described by the ball, the data and the cell. Such a statement can be proved by a sum-of-squares certificate \cite{lasserre2001,putinar1993,parrilo2003}, and the certificate can be verified in exact rational arithmetic, so that no solver tolerance enters the conclusion \cite{peyrl2008}. \ref{app:cert} describes the construction used for the plate. It is expensive compared with the precision analysis, and it is carried out for a fixed discretisation of the structure and of the law; it is nevertheless the only part of the chain that turns ``the prediction is defined for every admitted law'' from an expectation into a proved statement.
\section{Algorithms}\label{sec:alg}

\begin{figure}[tp]
\centering
\includegraphics[width=\textwidth]{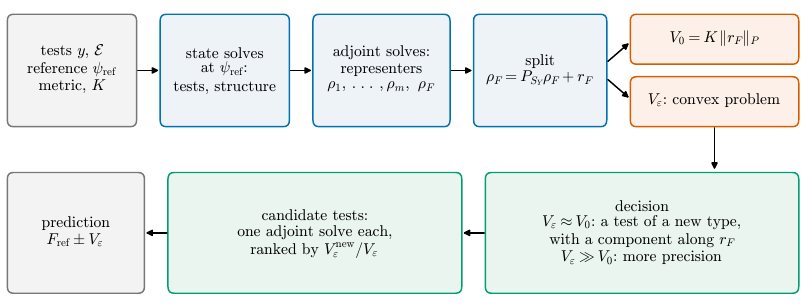}
\caption{The CFM procedure for one quantity of interest. One state solve and one adjoint solve per test and for the structure give the representers; the split into seen and hidden parts gives the two half-widths and the decision between a new type of test and more precision.}
\label{fig:flow}
\end{figure}

Figure~\ref{fig:flow} summarises the procedure for one quantity of interest, and Algorithms~\ref{alg:lin} and \ref{alg:nl} give its main computational parts. The constitutive function space is discretised by a basis $\{\phi_k\}_{k=1}^{n}$ with Gram matrix $P_{kl}=\ipP{\phi_k}{\phi_l}$; a functional with coefficient vector $\ell$ has the representer coefficients $P^{-1}\ell$, and all inner products of representers are evaluated through a Cholesky factor of $P$. The basis is a computational approximation of $\X$ and not a parametrisation of the law. It is chosen to resolve the states visited by the tests and by the structure (Section~\ref{sec:repr}), and its adequacy is checked by refinement, as a finite-element mesh is. When the order condition of Section~\ref{sec:class} holds, Theorem~\ref{thm:tol} reduces the linear problem exactly to the span of the representers, which removes the discretisation error of the linear layer altogether.

\begin{algorithm}[tp]
\caption{Linear CFM for one quantity of interest}\label{alg:lin}
\begin{algorithmic}[1]
\Require reference law $\psir$, metric $P$, radius $K$; test models with data $y$ and tolerances; structure model with quantity of interest $F$
\State solve the $m$ test problems and the structure problem at $\psir$; evaluate $\Ob[\psir]$, $\Fr$, $b=y-\Ob[\psir]$
\State one adjoint solve \eqref{eq:adjoint} per observation and for $F$: coefficient vectors $\ell_1,\dots,\ell_m,\ell_F$
\State representers $\rho_i=P^{-1}\ell_i$, $\rhoF=P^{-1}\ell_F$
\State orthogonal factorisation of $[\rho_1,\dots,\rho_m]$ in the $P$ inner product; $\rF=\rhoF-\PSY\rhoF$
\State $V_0=K\nP{\rF}$, $\sin\theta=\nP{\rF}/\nP{\rhoF}$
\State solve the convex problem \eqref{eq:dual} in $\mu\in\R^m$ (with \eqref{eq:admdual} if admissibility can be active): $V_\varepsilon$ and multipliers $\mu$
\State \Return $\Fr$, $V_0$, $V_\varepsilon$, $\rF$, $\mu$
\end{algorithmic}
\end{algorithm}

\begin{algorithm}[tp]
\caption{Finite-amplitude end point}\label{alg:nl}
\begin{algorithmic}[1]
\Require output of Algorithm~\ref{alg:lin}; sign $s=\pm1$
\State initialise $\psi\gets\psir+s\,v_{\mathrm{lin}}$ with the linear maximiser $v_{\mathrm{lin}}$ of \eqref{eq:dual}
\Repeat
\State solve the test and structure problems at $\psi$; adjoint solves; representers $\rho_i[\psi]$, $\rho_f[\psi]$
\State solve the quadratic programme for the step: linearised objective and constraints of \eqref{eq:nlp} with a Hessian approximation of the Lagrangian, restricted to $\spn\{\rho_f[\psi],\rho_i[\psi],\psi-\psir\}$
\State update $\psi$ and the multipliers $\alpha,\beta$ (line search on a merit function)
\Until the residual of \eqref{eq:kkt} and the constraint violations are below tolerance
\State check feasibility of $\psi$ directly; evaluate the margin $c$ of \eqref{eq:sosc}
\State \Return end-point law $\psi^\ast$, value $f[\psi^\ast]$, multipliers, margin $c$
\end{algorithmic}
\end{algorithm}

Candidate tests (Section~\ref{sec:acq}) are screened by one state solve and one adjoint solve each at $\psir$, which give $\rho_c$ and its hidden part; $\cos^2(\rF,r_c)$ gives the exact-data contraction \eqref{eq:cos}, and \eqref{eq:dual} with the row $\rho_c$ added gives $V_\varepsilon^{\mathrm{new}}$, by which the candidates are ranked.

\paragraph{Cost} Algorithm~\ref{alg:lin} requires $m+1$ state solves and $m+1$ adjoint solves, the latter with tangent matrices that are already factorised, and one convex problem in $m$ unknowns. Its cost is independent of the dimension $n$ of the constitutive discretisation apart from linear algebra with $P$. Each iteration of Algorithm~\ref{alg:nl} has the cost of Algorithm~\ref{alg:lin}. The screening adds one state and one adjoint solve per candidate. No sampling of the constitutive space is involved at any stage; Section~\ref{sec:cost} compares measured costs with those of a representation-based and a sampling-based treatment of the same problem.
\section{Numerical examples}\label{sec:examples}

\subsection{Two-element bar}\label{sec:toy}

\begin{figure}[tp]
\centering
\includegraphics[width=\textwidth]{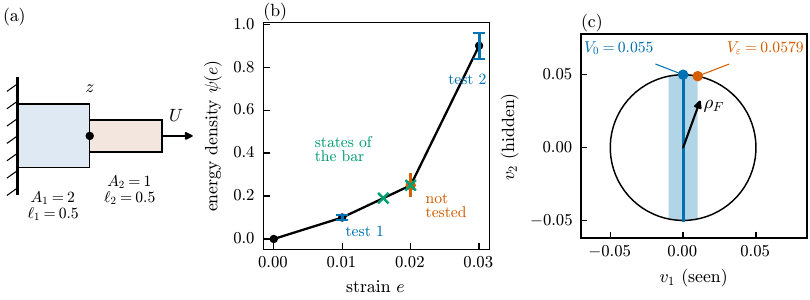}
\caption{Two-element bar. (a) Model. (b) Reference law, the two tests with their tolerances, the untested nodal value with the prior interval, and the states of the two elements at equilibrium. (c) Section of the linearised fibre in the plane of $v_1$ (seen by test~1) and $v_2$ (hidden): for exact data the fibre is the vertical segment and the half-width is $V_0$; with tolerance it is the shaded strip of the disk and the half-width is $V_\varepsilon$.}
\label{fig:toy}
\end{figure}

The first example can be followed by hand and shows every step of Algorithm~\ref{alg:lin}. A bar of two elements with lengths $\ell_1=\ell_2=0.5$ and cross-sections $A_1=2$, $A_2=1$ is fixed at one end and displaced by $U=0.018$ at the other (Fig.~\ref{fig:toy}a). The strain-energy density is piecewise linear on the strain nodes $e=0,\,0.01,\,0.02,\,0.03$ with $\psi(0)=0$, so that a law is the vector of its nodal values $\psi=(\psi_1,\psi_2,\psi_3)$; the metric is the Euclidean norm of this vector. The reference law is $\psir=(0.10,0.25,0.90)$ and the radius is $K=0.05$. Two tests measure the energy density directly, $\psi_1=0.10\pm0.01$ and $\psi_3=0.90\pm0.06$; the state $e=0.02$ is not tested. The quantity of interest is the stored energy $F[\psi]=\min_z\Pi(z;\psi)$, with $z$ the displacement of the interior node.

\emph{State solve.} The minimiser is $z^\ast=0.008$, with strains $e=(0.016,0.020)$ in the two elements, and $\Fr=0.315$. \emph{Sensitivities.} Since $F$ is a minimum, its derivative is the occupation of the constitutive states by the minimiser: element~1 contributes its volume $1$ with weights $0.4$ and $0.6$ to the nodes $0.01$ and $0.02$, element~2 its volume $0.5$ to the node $0.02$, so $\rhoF=(0.4,1.1,0)$. The observations have $\rho_1=(1,0,0)$ and $\rho_2=(0,0,1)$. \emph{Split.} $\SY$ is spanned by the first and third coordinate directions, and the hidden part is $\rF=(0,1.1,0)$. \emph{Half-widths.} $V_0=K\nP{\rF}=0.055$, and from \eqref{eq:dual} with the box tolerances, $V_\varepsilon=0.4\times0.01+1.1\sqrt{K^2-0.01^2}=0.0579$ (Fig.~\ref{fig:toy}c). The prediction is $F=0.315\pm0.058$, and $95\%$ of the half-width is due to the state that no test visits.

\emph{Decision.} Since $V_\varepsilon\approx V_0$, precision is not the issue: halving the tolerance of test~1 gives $0.0567$. A test of a new type, measuring $\psi_2$ with tolerance $0.02$, has the representer $(0,1,0)$, which is aligned with $\rF$; it gives $V_0=0$ and $V_\varepsilon=0.4\times0.01+1.1\times0.02=0.026$, and with tolerance $0.01$, $0.015$. In this example the equilibrium state is the same for all laws of the fibre, because the second element sits at a kink of the energy, so $F$ is exactly linear on the fibre and the linear values are exact.

\subsection{Phase-field bar with unknown degradation function}\label{sec:pf}

\subsubsection{Problem}

The second example is a bar in the AT2 phase-field model of fracture \cite{bourdin2000,pham2011,msekh2015}, with energy
\begin{equation}\label{eq:pfenergy}
\Pi(u,d;g)=\int_0^L\Big[\tfrac12\,g(d)\,E(x)\,u'^2+\frac{G_c}{2\ell}\big(d^2+\ell^2d'^2\big)\Big]\dd x,
\end{equation}
with $u(0)=0$, $u(L)=U$, $L=1$, $G_c=1$, $\ell=0.1$ and a stiffness $E(x)=1-0.02\exp[-((x-0.5)/0.1)^2]$ with a small imperfection that localises the damage at mid-span. Displacement and damage are discretised by 100 linear elements. The unknown constitutive function is the degradation function $g(d)$ on $[0,1]$. Its choice is known to control the stability of the homogeneous and localised solutions \cite{pham2011b,wu2017}, and it is not measured directly: a tension test gives the reaction as a function of the applied displacement.

The constitutive class is $g=g_{\mathrm{ref}}+v$ with $v$ in $\X=\{v\in H^4(0,1):v=v'=0\text{ at }d=0,1\}$ and $\nP{v}=\|v''''\|_{L^2}$, so that $g(0)=1$, $g(1)=0$ and the end slopes are fixed; the order $s=4$ satisfies the condition of Section~\ref{sec:class} for the second derivative $g''$ that enters the stability analysis. The reference $g_{\mathrm{ref}}$ is the quadratic $(1-d)^2$ plus a smooth perturbation in $\X$ of maximum size $8.3\e{-3}$ and norm $12.9$ (Fig.~\ref{fig:pfhidden}a), so that the example is not tied to the quadratic law. The radius is $K=10$.

\begin{figure}[tp]
\centering
\includegraphics[width=\textwidth]{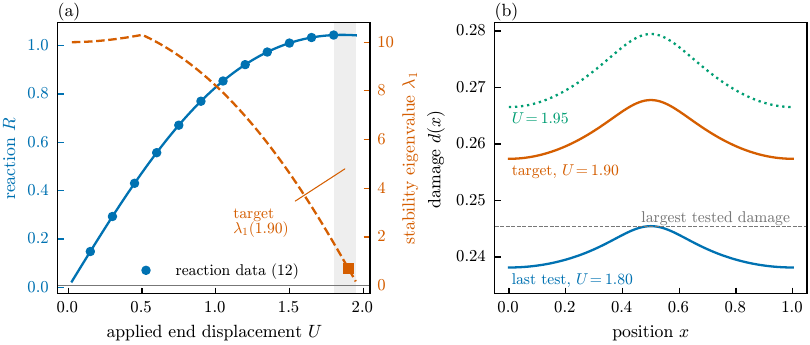}
\caption{Phase-field bar. (a) Reaction and stability eigenvalue along the loading path at the reference law, with the twelve reaction data and the target. (b) Damage profiles at the last test, at the target load and at $U=1.95$.}
\label{fig:pfsetup}
\end{figure}

The data are $m=12$ reactions at $U=0.15,0.30,\dots,1.80$, generated with $g_{\mathrm{ref}}$ and therefore nominal, with a joint Euclidean tolerance $\varepsilon=3\e{-5}$; the reactions are of order one, and the influence of the tolerance is examined below. The largest damage reached in the tests is $0.2455$. The quantity of interest is the smallest eigenvalue $\lambda_1$ of the reduced Hessian $S=J_{dd}-J_{du}J_{uu}^{-1}J_{ud}$ of \eqref{eq:pfenergy} at $U=1.90$, in the $L^2$ metric of the damage field. It measures the distance of the equilibrium from the loss of stability, which occurs near $U=1.97$ (Fig.~\ref{fig:pfsetup}a). At the target load the damage lies in $[0.2574,0.2678]$, just beyond the tested states (Fig.~\ref{fig:pfsetup}b). The prediction thus extrapolates slightly in load and in constitutive state, as every prediction of a failure load does.

\subsubsection{Hidden direction and half-widths}

\begin{figure}[tp]
\centering
\includegraphics[width=\textwidth]{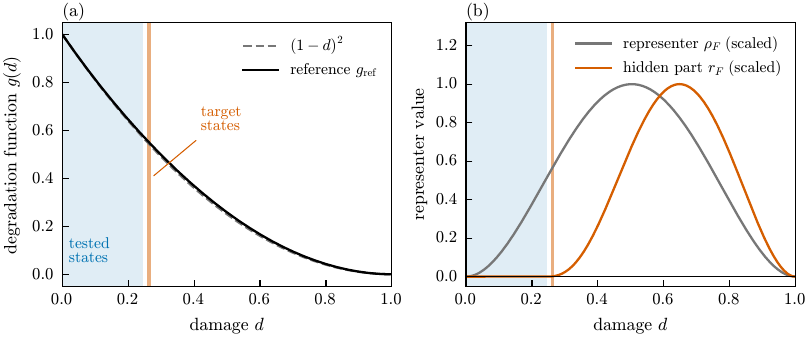}
\caption{Phase-field bar. (a) Reference degradation function, the states visited by the tests and the states of the target. (b) Representer $\rhoF$ of the stability eigenvalue and its hidden part $\rF$ as functions of the damage, each scaled to unit maximum. The hidden part vanishes to plotting accuracy on the tested states.}
\label{fig:pfhidden}
\end{figure}

Algorithm~\ref{alg:lin} gives $\Fr=\lambda_1(1.90)=0.700$ and $\nP{\rhoF}=0.1956$, so that the prior alone would allow $\lambda_1=0.700\pm1.96$, which is no prediction. The hidden part has $\nP{\rF}=6.12\e{-3}$:
\begin{equation}
\sin\theta=0.031,\qquad V_0=0.0612,\qquad V_\varepsilon=0.0874 .
\end{equation}
Only $3.1\%$ of the norm of the sensitivity of the eigenvalue is hidden from the twelve reactions, and this part produces $70\%$ of the uncertainty of the prediction, $\lambda_1(1.90)=0.700\pm0.087$. Figure~\ref{fig:pfhidden}b shows why: the hidden part is a perturbation of $g$ that vanishes on the tested states and is free beyond them, which includes the states of the target. The value of \eqref{eq:dual} computed in the span of the 13 representers was reproduced to ten digits in a space of discontinuous piecewise cubics for $v''''$ of dimension 10076, with breakpoints at all states visited by the model. The norm $\nP{\rF}$ changes by $0.05\%$ between meshes of 100, 200 and 400 elements. The derivative of $\lambda_1$ along $\rF$ agrees with central differences of full nonlinear solves to $7\e{-8}$ relative, and the reactions do not change along it to $10^{-13}$.

Table~\ref{tab:pfeps} shows the half-width as a function of the tolerance. With reactions measured to four digits the half-width is $0.30$; it decreases slowly and approaches $V_0=0.061$, which no precision of the reactions can reduce. For the value $\varepsilon=3\e{-5}$ used in the following both contributions are present. Monotonicity and positivity of $g$ were checked with Lemma~\ref{lem:activity} and are redundant on this fibre, with $\max_tV_\varepsilon(\gamma_t)/m_t=0.18$; they would become active for $K\approx56$.

\begin{table}[tp]
\centering\small
\caption{Phase-field bar: half-width of $\lambda_1(1.90)=0.700$ as a function of the tolerance of the reaction data ($K=10$).}
\label{tab:pfeps}
\begin{adjustbox}{max width=\textwidth}
\begin{tabular}{lcccccccc}
\toprule
$\varepsilon$ & $10^{-2}$ & $10^{-3}$ & $10^{-4}$ & $3\e{-5}$ & $10^{-5}$ & $3\e{-6}$ & $10^{-6}$ & $0$\\
$\varepsilon/\|y\|$ & $3.7\e{-3}$ & $3.7\e{-4}$ & $3.7\e{-5}$ & $1.1\e{-5}$ & $3.7\e{-6}$ & $1.1\e{-6}$ & $3.7\e{-7}$ & $0$\\
\midrule
$V_\varepsilon$ & 0.774 & 0.304 & 0.123 & 0.0874 & 0.0721 & 0.0650 & 0.0626 & 0.0612\\
\bottomrule
\end{tabular}
\end{adjustbox}
\end{table}

\subsubsection{Finite amplitude}

\begin{figure}[tp]
\centering
\includegraphics[width=\textwidth]{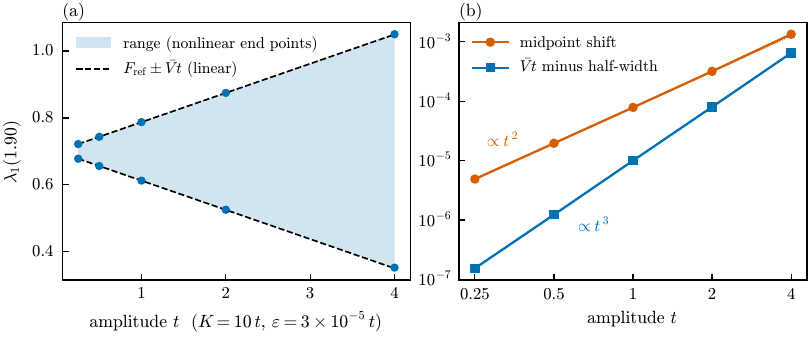}
\caption{Phase-field bar, finite amplitude. (a) Range of $\lambda_1(1.90)$ from the nonlinear end points and from the linear half-width. (b) Midpoint shift and defect of the half-width against the amplitude, with the slopes $2$ and $3$ of Theorem~\ref{thm:parity}.}
\label{fig:pfparity}
\end{figure}

The end points of the range were computed with Algorithm~\ref{alg:nl}. At $(K,\varepsilon)=(10,3\e{-5})$ the extreme laws give $\lambda_1-\Fr=+8.7498\e{-2}$ and $-8.7341\e{-2}$, each in four iterations, with both constraints active to $10^{-8}$; the half-width is $0.99989\,V_\varepsilon$ and the midpoint is shifted by $7.8\e{-5}$. The second-order check \eqref{eq:sosc} at the upper end point gives $\alpha=6.7757\e{-4}$ and a margin $c=6.7727\e{-4}>0$; second differences of the Lagrangian from state solves are negative in all sampled tangent directions. Table~\ref{tab:pfparity} and Fig.~\ref{fig:pfparity} show the continuation in the amplitude $t$ of Theorem~\ref{thm:parity}. Over a factor 16 in amplitude, up to a range in which $\lambda_1$ varies between $0.35$ and $1.05$, the half-width stays within $0.2\%$ of the linear value; its defect grows by a factor $8$ and the midpoint shift by a factor $4$ per doubling of $t$, the latter with a remainder that grows by $16$, as \eqref{eq:parity} predicts.

\begin{table}[tp]
\centering\small
\caption{Phase-field bar: finite-amplitude end points for $(K,\varepsilon)=t\,(10,3\e{-5})$. The last column is the leading term $s_2t^2$ of \eqref{eq:parity}, with $s_2$ computed at the reference.}
\label{tab:pfparity}
\begin{adjustbox}{max width=\textwidth}
\begin{tabular}{cccccc}
\toprule
$t$ & $\bar Vt$ & $f^+-\Fr$ & $f^--\Fr$ & half-width$/\bar Vt$ & midpoint shift ($s_2t^2$)\\
\midrule
0.25 & $2.1857\e{-2}$ & $+2.18620\e{-2}$ & $-2.18523\e{-2}$ & 0.999993 & $4.875\e{-6}$ ($4.874\e{-6}$)\\
0.5 & $4.3715\e{-2}$ & $+4.37328\e{-2}$ & $-4.36938\e{-2}$ & 0.999972 & $1.952\e{-5}$ ($1.950\e{-5}$)\\
1 & $8.7429\e{-2}$ & $+8.74975\e{-2}$ & $-8.73408\e{-2}$ & 0.999886 & $7.834\e{-5}$ ($7.799\e{-5}$)\\
2 & $1.7486\e{-1}$ & $+1.75096\e{-1}$ & $-1.74461\e{-1}$ & 0.999543 & $3.176\e{-4}$ ($3.120\e{-4}$)\\
4 & $3.4972\e{-1}$ & $+3.50408\e{-1}$ & $-3.47728\e{-1}$ & 0.998145 & $1.340\e{-3}$ ($1.248\e{-3}$)\\
\bottomrule
\end{tabular}
\end{adjustbox}
\end{table}

\paragraph{Several quantities of interest} For the three quantities $\lambda_1(1.90)$, the reaction $R(1.90)$ and $\lambda_1(1.95)$, the hidden fractions are $0.031$, $0.0027$ and $0.045$. The normalised hidden Gram matrix \eqref{eq:Ghid} has the eigenvalues $2.97\e{-3}$, $2.5\e{-5}$ and $6.5\e{-8}$: it has numerical rank one, and a single hidden constitutive direction carries the uncertainty of all three predictions. This is favourable for test design, since one well-chosen test serves all three. With tolerance, the ratio of $V_\varepsilon(a)$ to the exact-data ellipsoid value $K(a\T G_{\mathrm{hid}}a)^{1/2}$ varies between $1.12$ and $3.07$ over $34$ directions, so the hidden Gram matrix alone does not describe the image of the fibre at $\varepsilon>0$ (Section~\ref{sec:linear}). Nonlinear end points in eight directions of the plane of the first two quantities agree with the linear support values within $0.8\%$.

\subsubsection{Representations}

\begin{figure}[tp]
\centering
\includegraphics[width=\textwidth]{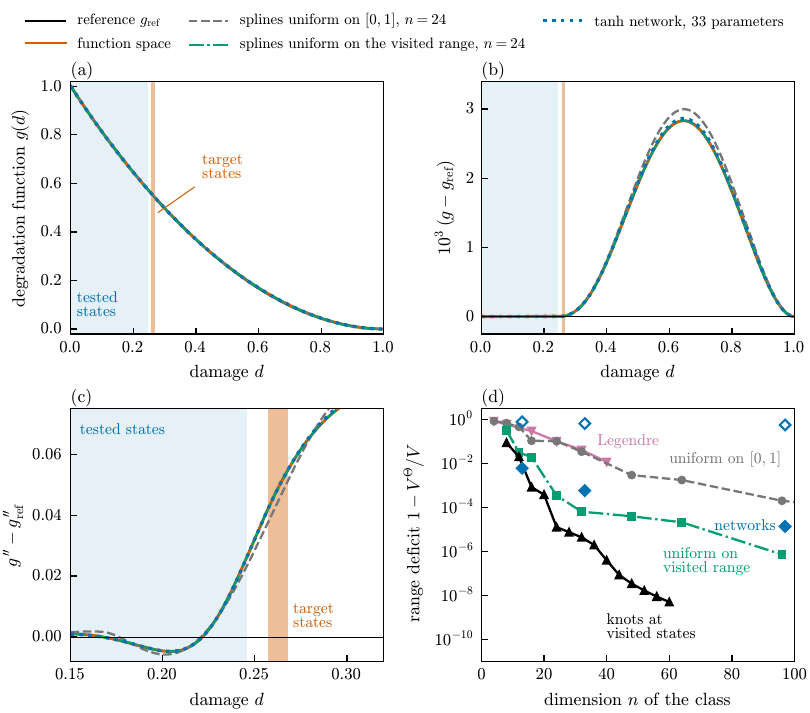}
\caption{Phase-field bar: extreme laws and representations. (a) Reference degradation function and the upper extreme law for $\lambda_1(1.90)$, computed in the function space, in two spline spaces of dimension 24 and with a tanh network of 33 parameters; at this scale the laws coincide. (b) Their deviation from the reference. (c) Deviation of the second derivative, through which the stability eigenvalue sees the law, near the tested and the target states. (d) Deficit of the range that a class of dimension $n$ can express. Linear classes: exact values from Proposition~\ref{prop:inner}. Networks: tanh networks with two hidden layers of width 2, 4 and 8; open diamonds are the result of a constrained search started from the reference, filled diamonds the best member found by projecting the function-space extreme law onto the network and refining it locally.}
\label{fig:pfrepr}
\end{figure}

Figure~\ref{fig:pfrepr}a shows the upper extreme law, the degradation function that maximises $\lambda_1(1.90)$ over the fibre. It differs from the reference by at most $2.8\e{-3}$ (Fig.~\ref{fig:pfrepr}b). The deviation vanishes on the tested states and is largest near $d=0.65$, where nothing is measured: it is the smoothest deviation in the declared metric that changes the curvature of $g$ at the target states while leaving the tested states unchanged. The eigenvalue sees the law through $g''$ at the target states (Fig.~\ref{fig:pfrepr}c). There the extreme laws computed with splines of dimension 24 on the visited range and with a network of 33 parameters coincide with the function-space law, while splines of the same dimension that are uniform on $[0,1]$ reach $91\%$ of its curvature deviation and $90\%$ of the range.

Figure~\ref{fig:pfrepr}d and Table~\ref{tab:pfrepr} show how much of the range classes of given dimension can express (Proposition~\ref{prop:inner}). The classes are piecewise polynomial or polynomial representations of $v''''$. Read as discretisations of $\X$, the results show how fast a basis converges; read as constitutive models of fixed dimension, they show how much of the range such a model hides. With 24 unknowns, Legendre polynomials and splines that are uniform on $[0,1]$ miss $10\%$ of the range; splines that are uniform on the visited range of damage miss $0.03\%$; and splines with knots selected among the edges and midpoints of the bands of damage visited at the test and target loads miss $0.001\%$. Knots above the visited range contribute nothing. The classes differ by four orders of magnitude at equal dimension, and the difference is explained entirely by whether the class resolves the states at which the representers have their knots. The same ordering holds at finite amplitude: the nonlinear end points computed within each class reproduce the linear deficits (Legendre 24: $10.0\%$; visited-range splines, 24 unknowns: $0.035\%$).

\begin{table}[tp]
\centering\small
\caption{Phase-field bar: range deficit $1-V_\varepsilon^\Theta/V_\varepsilon$ of linear classes of dimension $n$. In the last column the knots are selected greedily among the edges and midpoints of the visited bands by the deficit itself, which gives an upper envelope of what knot placement can achieve; the rule that places knots at the edges and means of the visited bands without reference to the target gives $3.0\e{-7}$ with $n=60$.}
\label{tab:pfrepr}
\begin{adjustbox}{max width=\textwidth}
\begin{tabular}{ccccc}
\toprule
$n$ & Legendre & splines, uniform on $[0,1]$ & splines, visited range & splines, knots at visited states\\
\midrule
16 & $2.9\e{-1}$ & $1.1\e{-1}$ & $1.9\e{-2}$ & $8.2\e{-4}$\\
24 & $1.0\e{-1}$ & $1.0\e{-1}$ & $3.4\e{-4}$ & $1.3\e{-5}$\\
32 & $4.1\e{-2}$ & $3.4\e{-2}$ & $6.3\e{-5}$ & $4.3\e{-6}$\\
48 & -- & $2.9\e{-3}$ & $3.9\e{-5}$ & $3.4\e{-8}$\\
\bottomrule
\end{tabular}
\end{adjustbox}
\end{table}

For neural networks two effects must be distinguished. Small tanh networks (13, 33 and 97 parameters) contain members that reach $99.4\%$, $99.94\%$ and $99.999\%$ of the upper end point; these were found by projecting the function-space extreme law onto the network and refining the projection locally, and they are feasible with the norm evaluated by a certified quadrature. A constrained search over the network parameters started from the reference, which is what one would do without the function-space solution, reached $24\%$, $36\%$ and $45\%$ of the range after 400 iterations. Expressivity is therefore not the limitation of these networks; the search is. In both cases the reported range is an inner approximation, and its distance from the true range is known only because the function-space problem has been solved. The value-loss relation \eqref{eq:valueloss} was also confirmed: a relative error of $1\%$ of the extreme law in the metric costs about $0.004\%$ of the end-point value, so that agreement of end-point values is a weak test of a learned law.

\subsubsection{Selecting the next test}

\begin{figure}[tp]
\centering
\includegraphics{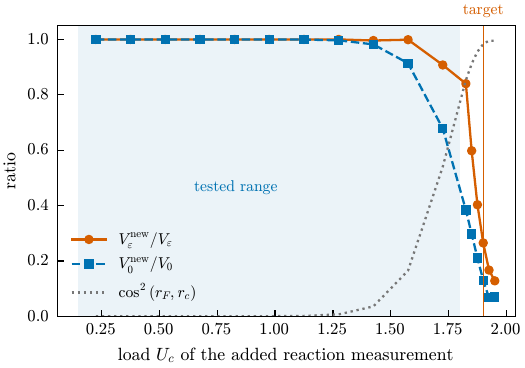}
\caption{Phase-field bar: effect of one additional reaction measured at load $U_c$ on the half-width of $\lambda_1(1.90)$, for exact data and with tolerance, and the alignment \eqref{eq:cos} of the candidate with the hidden part of the target.}
\label{fig:pfacq}
\end{figure}

The screening of Section~\ref{sec:alg} was applied to 17 candidate reaction measurements and three candidate stability measurements (Fig.~\ref{fig:pfacq}). Reactions at loads up to $1.575$, between the existing test loads, reduce $V_\varepsilon$ by less than $0.5\%$; one at $1.725$, between the last two test loads, reduces it by $9\%$. Reactions beyond the tested range are decisive: one reaction at $U=1.95$ has $\cos^2=0.995$ and reduces the half-width from $0.0874$ to $0.0112$, that is, to $12.8\%$. The prediction was verified at finite amplitude: with the added row, the nonlinear end points give a half-width of $0.128034$ of the original one, against the linear value $0.128019$. For the three quantities of interest together, the same test reduces the half-widths to $13\%$, $18\%$ and $22\%$.

Table~\ref{tab:pffloor} relates this to Proposition~\ref{prop:limits}. With reactions restricted to $U\le1.80$, more readings lower $\nP{\rF}$ towards the coverage floor $1.5\e{-3}$, which is not zero because the tests do not reach the target states; at the same time, with a fixed tolerance per reading, the half-width stops decreasing after about 56 readings. Extending the range lowers the floor, and it vanishes when the tests reach $U=1.90$. The remaining half-width is then limited by precision alone.

\begin{table}[tp]
\centering\small
\caption{Phase-field bar: reactions at equally spaced loads up to $U_{\max}$. Coverage floor of Proposition~\ref{prop:limits}(iii), $\nP{\rF}$ for $n\approx12$ and $n\approx170$ readings, and limit of the half-width for a fixed tolerance $\varepsilon/\sqrt{12}$ per reading.}
\label{tab:pffloor}
\begin{adjustbox}{max width=\textwidth}
\begin{tabular}{cccccc}
\toprule
$U_{\max}$ & largest tested damage & coverage floor & $\nP{\rF}$, $n\approx12$ & $\nP{\rF}$, $n\approx170$ & $V$ (many readings)\\
\midrule
1.80 & 0.2455 & $1.5\e{-3}$ & $6.1\e{-3}$ & $2.0\e{-3}$ & 0.064\\
1.85 & 0.2565 & $1.5\e{-4}$ & $1.8\e{-3}$ & $3.7\e{-4}$ & 0.035\\
1.88 & 0.2632 & $3.8\e{-6}$ & $1.2\e{-3}$ & $4.3\e{-5}$ & 0.020\\
1.90 & 0.2678 & 0 & $7.9\e{-4}$ & $4.0\e{-5}$ & 0.011\\
\bottomrule
\end{tabular}
\end{adjustbox}
\end{table}

\subsubsection{Cost}\label{sec:cost}

Table~\ref{tab:pfcost} lists the measured cost of answering the same question, the range of $\lambda_1(1.90)$ over the laws consistent with the data, by different routes on one processor core. All routes were implemented by us on the same model; the comparison is not a benchmark of published codes. The function-space computation costs seconds. A constrained optimisation in a class of 24 Legendre polynomials costs about as much as the nonlinear CFM end points and returns $90\%$ of the range. Network searches cost one to two orders of magnitude more and return less than half. Rejection sampling is not feasible: of $1000$ laws drawn uniformly from the ball of the 24-term class none reproduced the data within tolerance (the smallest misfit was $9.2\,\varepsilon$), and the acceptance rate estimated from $4\e{7}$ samples of the linearised map is $6\e{-7}$, about $1.6\e{6}$ state solves per accepted sample.

\begin{table}[tp]
\centering\small
\caption{Phase-field bar: cost of computing the range of $\lambda_1(1.90)$ (one core; one state solve along the loading path takes $0.31$~s).}
\label{tab:pfcost}
\begin{adjustbox}{max width=\textwidth}
\begin{tabular}{lcc}
\toprule
Route & time (s) & fraction of the range found\\
\midrule
CFM, linear (Algorithm~\ref{alg:lin}) & 5.3 & 1 (linear value)\\
CFM, both nonlinear end points (Algorithm~\ref{alg:nl}) & 37 & 1\\
CFM, screening of 20 candidates for 3 targets & 15 & --\\
Constrained optimisation, Legendre class, $n=24$, both ends & 35 & 0.90\\
Network search, width 2 / 4 / 8, per end & 1640 / 2250 / 5040 & 0.24 / 0.36 / 0.45\\
Rejection sampling, 1000 samples & 489 & none accepted\\
\bottomrule
\end{tabular}
\end{adjustbox}
\end{table}

\subsection{Open-hole composite lamina}\label{sec:plate}

\subsubsection{Problem}

\begin{figure}[tp]
\centering
\includegraphics[width=\textwidth]{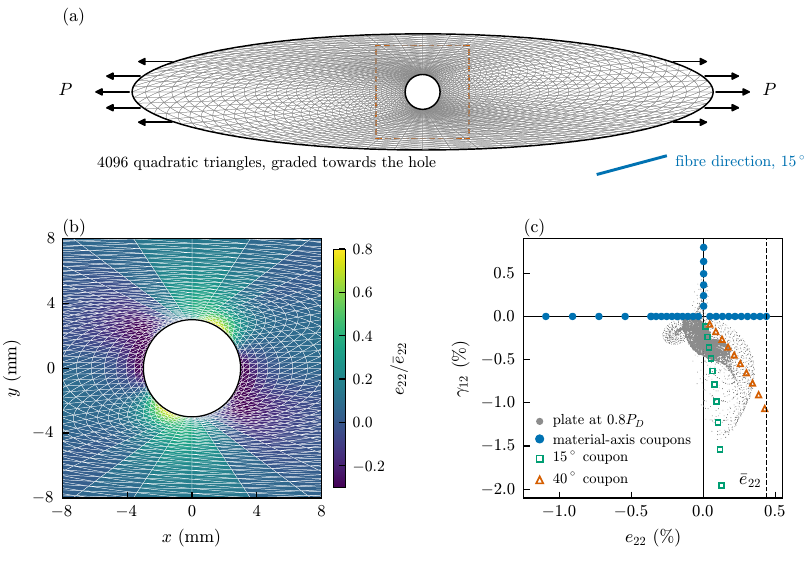}
\caption{Open-hole lamina: model and discretisation. (a) Planform, mesh, loading and fibre direction; the dashed square marks the region shown in (b). (b) Mesh near the hole, coloured by the largest transverse strain of each element relative to its bound $\bar e_{22}=0.436\%$, at $0.8P_D$ with the law $\psi^\dagger$. (c) Constitutive states in the plane of transverse and shear strain: the states of the plate at the quadrature points, the states of the four material-axis coupons, and those of off-axis coupons at $15^\circ$ and $40^\circ$; the material-axis coupons continue along the axes to the boundary of $\Omega_c$.}
\label{fig:platemodel}
\end{figure}

The third example is two-dimensional and uses a constitutive function of three variables. A unidirectional AS4/3501-6 lamina in plane stress has an elliptical planform with semi-axes $50$~mm and $10$~mm, thickness $1$~mm and a central hole of radius $3$~mm; the fibres make an angle of $15^\circ$ with the long axis (Fig.~\ref{fig:platemodel}a). The boundary tractions are those of a homogeneous uniaxial stress along the long axis with resultant $P$, so that the plate without hole would be homogeneous for every hyperelastic law. The plate is discretised by 4096 isoparametric quadratic triangles on a mapped mesh with 64 divisions around the hole and 32 radially, graded towards the hole (8320 nodes); stiffness and residual are integrated with a seven-point rule of degree five, and three displacement components at the hole are fixed to remove the rigid motions of the balanced load. The strains at the 28\,672 quadrature points and at nine points per element edge on the hole are the constitutive states of the structure (Fig.~\ref{fig:platemodel}b,c).

The unknown is the strain-energy density $\psi(e_{11},e_{22},\gamma_{12})$ in material axes on the strain domain $\Omega_c=[-1.175,1.38]\%\times[-2.0,0.436]\%\times[-2,2]\%$, the box of the uniaxial failure strains of the material in the first World-Wide Failure Exercise \cite{soden1998}. Admissibility requires evenness in the shear strain, a natural state at zero strain and convexity on $\Omega_c$. The metric is the Sobolev norm of order five of the dimensionless energy in strains scaled by the box, with unit weights; order five is the smallest integer order for which the third derivatives entering the sensitivity of tangent quantities are bounded in three dimensions. For the computation the function space is discretised by polynomials of total degree at most 14 in the scaled strains, a space of dimension 680 in which the natural state is imposed by four linear conditions and evenness by symmetrisation. This is a discretisation of the Sobolev class, not a polynomial model of the material. Since the discretised fibre is contained in the fibre, the computed widths are lower bounds; their change under refinement is reported below.

The reference $\psir$ is the linear orthotropic law with the moduli of \cite{soden1998}. To generate data and to choose the radius we use a nonlinear law $\psi^\dagger$ fitted to the shear and transverse-compression curves of \cite{soden1998}, whose nonlinear part amounts to $40\%$ of the shear strain at failure. Its distance from the reference is $\nP{\psi^\dagger-\psir}=64.156$, of which $98\%$ comes from the fifth derivatives, and we set $K$ to this value, the smallest radius for which the prior contains $\psi^\dagger$. The test programme consists of four coupons in material axes, shear, transverse tension, transverse compression and axial tension, each with a strain rosette and ten load levels, generated with $\psi^\dagger$. This gives 120 readings of the stress at a measured strain, which are exactly linear in the law. The tolerance per reading is $0.5\%$ of the stress plus the stress of a strain error of $20$ microstrain at the reference stiffness; the strain error is thus converted into a stress tolerance at the measured state, and its effect on the state itself is not modelled. Figure~\ref{fig:platemodel}c shows that the coupon states lie on the coordinate axes of the strain space, while the states of the plate fill a two-dimensional region between them. The data are synthetic; the purpose of the example is the structure of the result, not a statement about the material. The load is referred to $P_D=1792$~N, at which the plate with law $\psi^\dagger$ first reaches the boundary of $\Omega_c$.

\subsubsection{Precision}

The fibre is non-empty, with $K_{\min}=16.5$ from Proposition~\ref{prop:gram}, and the plate is linearised at its minimum-norm element. Table~\ref{tab:plate} gives the first-order ranges of four quantities at $0.6P_D$. The secant compliance is predicted within a range of $6.2\%$ without and $3.2\%$ with the convexity condition, which is thus active in the sense of Lemma~\ref{lem:activity}. The mean transverse strain in the zone around the hole has a range larger than its value. The value for $\psi^\dagger$ lies inside every range, at all loads and for all quantities. Finite-amplitude end points for the compliance, computed with convex extreme laws followed along the load path, give $6.927\e{-5}$ and $7.153\e{-5}$~mm/N against the linear $6.925\e{-5}$ and $7.148\e{-5}$: the width is preserved and the midpoint shifts by about $3\%$ of the half-width. The data are not nominal here, so Theorem~\ref{thm:parity} does not apply directly, but the behaviour is the same.

\begin{table}[tp]
\centering\small
\caption{Open-hole lamina at $0.6P_D$: width of the first-order range in per cent of the value, and its dependence on the declared information (secant compliance, without convexity).}
\label{tab:plate}
\begin{adjustbox}{max width=\textwidth}
\begin{tabular}{lcc}
\toprule
Quantity of interest & without convexity & with convexity\\
\midrule
secant compliance & 6.19 & 3.18\\
mean shear strain, hole zone & 7.12 & 5.61\\
mean transverse strain, hole zone & 146 & 107\\
mean shear strain, far field & 5.27 & 4.71\\
\midrule
\multicolumn{3}{l}{secant compliance, without convexity, under changed declarations:}\\
\multicolumn{3}{l}{\quad radius $K/2$: 5.45;\quad $2K$: 6.69;\quad 40 levels per coupon: 6.09;\quad tolerance $0.1\%$: 4.45}\\
\multicolumn{3}{l}{\quad degree of the discretisation 10 / 14 / 16:\ \ 5.38 / 6.19 / 6.49}\\
\bottomrule
\end{tabular}
\end{adjustbox}
\end{table}

The lower part of Table~\ref{tab:plate} shows what limits the precision. Four times as many readings on the same coupons change the range of the compliance from $6.19\%$ to $6.09\%$: the uncertainty lies in strain states off the coupon paths, and repeating the paths does not reach it, as Proposition~\ref{prop:limits}(iii) states. The tolerance and the radius have a moderate influence. The discretisation has not converged at degree 14. Since the discretised fibre is contained in the fibre, the widths are lower bounds of the converged values, and we use them to compare designs, not as final values.

\subsubsection{Validity}

\begin{figure}[tp]
\centering
\includegraphics[width=\textwidth]{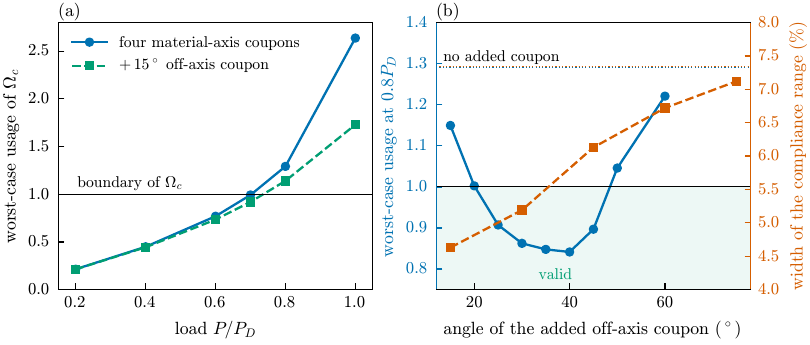}
\caption{Open-hole lamina: validity and acquisition (first order). (a) Worst-case usage of the strain domain over the fibre against the load; the prediction is valid where the usage is below one. (b) One additional off-axis coupon at $0.8P_D$: worst-case usage over all outcomes the centre law can produce, and width of the range of the secant compliance for the outcome generated by $\psi^\dagger$; dotted lines are the values without the added coupon.}
\label{fig:plate}
\end{figure}

By Proposition~\ref{prop:Eprior} the prior alone gives no validity, since $\nP{\psir}=23.0<K$; a first-order estimate places the limit of the prior set at $5\%$ of $P_D$. With the coupon data the situation changes. Figure~\ref{fig:plate}a shows the largest first-order usage of the strain domain over the fibre, that is, the largest ratio of a strain component at any point of the plate to its bound in $\Omega_c$. It reaches one at $0.70P_D$; the binding state is the transverse strain at the hole edge. Information extends the range of loads in which the prediction is defined by an order of magnitude.

At $0.8P_D$ the usage is $1.29$ and the prediction is not valid. Figure~\ref{fig:plate}b shows the effect of one additional off-axis tension coupon. For validity we use the criterion of Proposition~\ref{prop:Cpair} at first order: the largest usage over all laws of the fibre whose response in the new coupon lies within twice the tolerance of that of the centre law, which bounds the usage for every outcome that the centre law can produce. The $15^\circ$ coupon, which has the orientation of the plate itself, reduces the range of the compliance most, from $7.3\%$ to $4.6\%$, and leaves the usage at $1.15$. Coupons between $25^\circ$ and $45^\circ$ bring the usage below one, with the optimum near $40^\circ$ ($0.84$), and reduce the range of the compliance less. The reason is mechanical: the compliance is governed by the far field, an off-axis state at $15^\circ$, while the binding state at the hole edge depends on the law over the whole hole zone, where the hoop direction makes angles of about $30^\circ$ to $40^\circ$ with the fibres. Figure~\ref{fig:platemodel}c shows the same in the strain space: the states of the $15^\circ$ coupon run through the cluster of far-field states, and those of the $40^\circ$ coupon run along the branch of states that leads to the hole edge. The test that makes the prediction most precise and the test that makes it valid are different tests, as anticipated in Section~\ref{sec:validity}.

\subsubsection{Certified validity}

The usage values above are first-order estimates. To show that validity can be established rigorously, we certified the gate of Lemma~\ref{lem:gate} at $0.3P_D$ ($537.6$~N) for the finite-element model and the degree-14 law space, on a declared cell of the data-consistent set: the laws in the ball of radius $K$ that satisfy the 120 coupon inequalities and whose 16 leading amplitude coordinates lie in one box of a subdivision of the amplitude image into 2001 boxes, namely the box with the largest bound of the retained residual. Convexity is not used, so the set is an outer set of the corresponding part of the fibre. The cell is covered by eight parent boxes; the cover is verified by an independent computation in integer arithmetic. On each parent the gate is reduced to the non-negativity of a polynomial on a semi-algebraic set and proved by a sum-of-squares certificate that is accepted in exact rational arithmetic (\ref{app:cert}). Table~\ref{tab:cert} lists the result.

\begin{table}[tp]
\centering\small
\caption{Open-hole lamina: exact certificate of the gate at $0.3P_D$ on the eight parent boxes of the declared cell. A piece is a sub-box certified by one sum-of-squares identity; the margin is the smallest certified excess of the capacity term over the residual bound in the comparison \eqref{eq:closure}.}
\label{tab:cert}
\begin{adjustbox}{max width=\textwidth}
\begin{tabular}{ccc@{\qquad\qquad}ccc}
\toprule
Parent & pieces & smallest margin & Parent & pieces & smallest margin\\
\midrule
0 & 1 & $4.41\e{-6}$ & 4 & 7 & $7.34\e{-7}$\\
1 & 7 & $1.02\e{-6}$ & 5 & 1 & $3.68\e{-6}$\\
2 & 18 & $3.19\e{-6}$ & 6 & 7 & $4.52\e{-7}$\\
3 & 1 & $3.08\e{-5}$ & 7 & 1 & $3.60\e{-5}$\\
\bottomrule
\end{tabular}
\end{adjustbox}
\end{table}

All 43 pieces are accepted with positive margin. Parents 3, 5 and 7 are certified with the unshifted predictor; parents 0, 1, 2, 4 and 6 use a predictor shifted to the parent, and four of them had to be partitioned, since the certified residual bound reaches between about $98\%$ and $99.9\%$ of the capacity on their pieces. The statement proved is therefore the following: for every law of the declared cell, the finite-element problem at $0.3P_D$ has exactly one equilibrium in the ball around its predictor, and all its strains lie in $\Omega_c$. The statement is conditional on the model inputs that the certificate takes as given, namely the finite-element operators, the Gram matrix of the retained residual images, the capacity and the bounds of the truncated terms; these come from one computational chain and have not been reproduced by an independent implementation. It concerns one cell and the fixed discretisation; it is not a statement about the whole fibre, the full Sobolev class or the continuum problem. The remaining cells of the subdivision can be treated by the same procedure; this has not been done. We regard the result as a demonstration that the validity of a prediction over a set of constitutive laws is accessible to proof, and as the starting point of the function-space formulation discussed in Section~\ref{sec:discussion}.
\section{Discussion}\label{sec:discussion}

\paragraph{What is declared and what is computed} CFM does not remove assumptions; it makes them explicit and separates them from the mechanics. The declarations are the reference law, the metric, the radius $K$, the admissibility conditions and the error model of the tests. The subspace of hidden directions $\ker L$ is fixed by the mechanics of the tests alone. The size of the hidden part of a prediction depends in addition on the metric, and the half-width scales with $K$ in the hidden-dominated regime. The radius is therefore the declaration that deserves most scrutiny. Two checks are available within the method: the radius must not be smaller than $K_{\min}$ of Proposition~\ref{prop:gram}, and the sensitivity of the result to $K$ is directly computable (Table~\ref{tab:plate}). The plate example also shows what a credible radius means: a prior that contains a realistic nonlinear law is a bound on high derivatives, a regularity budget, and it is large when expressed as a norm.

\paragraph{Relation to Bayesian and learning-based treatments} The Gram form \eqref{eq:master} contains the linear Bayesian posterior for every fixed noise level (Section~\ref{sec:reading}); the two descriptions share the hidden directions and differ in what is reported, a bound over a set against a credible interval. The set-based statement needs no sampling, which matters when the acceptance region is as thin as in Section~\ref{sec:cost}, and it gives the decision between a new type of test and more precision in closed form; where a probability of exceedance is required, the Bayesian description is the appropriate one, and the representers computed here serve it as well. For learned constitutive laws, Sections~\ref{sec:repr} and \ref{sec:pf} show that the spread of a trained representation can under-report the range for two separate reasons, a class that does not resolve the constitutive states that matter and a search that does not find the extreme members the class contains. The first is cured by placing basis functions or features at the states visited by the tests and by the structure; for the second, the function-space end points provide the reference. A learned law remains a natural choice of reference law $\psir$, and CFM then quantifies what the data that trained it leave open.

\paragraph{Limits} Four limits of the present work should be stated. First, the finite-amplitude end points are attained values with a local optimality check; a rigorous outer bound of the range at finite amplitude requires a global argument, of the kind used for validity in \ref{app:cert}, and is not given here. Second, the certificate of validity concerns one cell of the data-consistent set of a fixed finite-element model and a fixed polynomial discretisation of the law space, and it rests on inputs from a single computational chain; its extension to all cells is a matter of computation, its extension to the full Sobolev class and to the continuum problem is a matter of analysis. Third, the plate example uses synthetic data, a radius calibrated on the law that generated them and a discretisation of the law space that has not converged; its numbers illustrate the structure of the result. Fourth, the theory assumes a regular solution branch. Irreversibility, contact and other inequality constraints on the state enter through active sets, which is compatible with the adjoint formula on a fixed active set but not treated here; the phase-field example is restricted to the reversible regime before localisation.

\paragraph{Outlook} Three extensions follow directly. The first is the function-space form of the validity certificate, in which the residual is bounded through functionals of the law that are bounded in the declared metric, so that no polynomial truncation is needed. The second is the treatment of the fracture problem proper: the degradation function and the dissipation function of a phase-field model as unknown constitutive functions, with the peak load and the critical energy release as quantities of interest and with the irreversibility constraint on the state. The third is the laminate, in which the fibre is built from lamina-level coupon data and the prediction concerns a laminate with a stress concentration, the setting in which the distinction between validity and precision found in Section~\ref{sec:plate} is of direct practical interest.

\section{Conclusions}\label{sec:conclusions}

A finite test programme and a declared regularity bound leave a set of constitutive functions, the constitutive fibre, and a structural prediction inherits a range from it. Constitutive-fibre mechanics analyses this set through the mechanics of the tests and of the structure. Adjoint solves represent observations and quantities of interest in the constitutive function space; the part of the sensitivity of a prediction that no observation sensitivity explains is hidden, and for exact data it alone determines the half-width. For data with tolerance the half-width is a finite convex problem whose cost does not depend on the constitutive discretisation, and the comparison of the two values separates what a more precise test can remove from what only a test of a new type can remove. Tests that do not visit the constitutive states of the prediction leave a floor that neither their number nor their precision reduces. At finite amplitude the extreme laws satisfy an optimality system with moving representers, and under nominal data the linear half-width is accurate to third order. A constitutive representation can only under-report the range, and the deficit is computable. Whether the prediction is defined for every admitted law is a separate property of the same set, governed by different constitutive directions and improved by different tests; for the composite plate it was proved on a cell of the set by a certificate accepted in exact arithmetic.

In the phase-field example, three per cent of the sensitivity of the stability eigenvalue was hidden from twelve accurate reaction measurements and produced seventy per cent of its uncertainty; one measurement beyond the tested range removed most of it. The constitutive information that a prediction needs is small and specific, and the mechanics identifies it.

\section*{CRediT authorship contribution statement}
\textbf{Altaf Ahmad Lone:} Conceptualization, Methodology, Software, Validation, Formal analysis, Investigation, Writing -- original draft, Visualization. \textbf{R. T. Durai Prabhakaran:} Supervision, Writing -- review \& editing. \textbf{Timon Rabczuk:} Conceptualization, Methodology, Supervision, Writing -- review \& editing. \textbf{Xiaoying Zhuang:} Supervision, Writing -- review \& editing.

\section*{Declaration of competing interest}
The authors declare that they have no known competing financial interests or personal relationships that could have appeared to influence the work reported in this paper.

\section*{Data availability}
The code and the data that reproduce the examples, including the certificate files and their replay scripts, will be made openly available on GitHub upon publication.

\appendix
\section{Proofs}\label{app:proofs}

\paragraph{Theorem~\ref{thm:exact}} For $\E=\{0\}$ and $b=0$ the linearised fibre is $\{v\in\ker L:\nP{v}\le K\}$. For $v\in\ker L$, $\ipP{\rho_a}{v}=\ipP{r_a}{v}\le\nP{r_a}\nP{v}$, with equality for $v=Kr_a/\nP{r_a}$, which is feasible. The set is symmetric, so the range is symmetric. The image of the ball of $\ker L$ under $v\mapsto(\ipP{r_{F_j}}{v})_j$ is the ellipsoid with support function $K\nP{\sum_ja_jr_{F_j}}=K(a\T G_{\mathrm{hid}}a)^{1/2}$. \qed

\paragraph{Theorem~\ref{thm:tol}} For $v\in C$ and any $\mu\in\R^m$,
\[
\ipP{\rho_a}{v}=\ipP{\rho_a-L^\ast\mu}{v}+\mu\T(Lv-b)+\mu\T b\le K\nP{\rho_a-L^\ast\mu}+h_\E(\mu)+\mu\T b,
\]
which is weak duality. The primal problem maximises a continuous linear functional over the intersection of a ball with a closed convex set, and $v=0$ is strictly feasible; by Fenchel--Rockafellar duality \cite{rockafellar1970} there is no gap and the dual minimum is attained. If $v=v_\parallel+v_\perp$ with $v_\parallel\in\spn\{\rho_a,\rho_1,\dots,\rho_m\}$ and $v_\perp$ orthogonal to it, then $v_\parallel$ has the same objective and observations and a norm not larger, so a maximiser exists in the span. For the Euclidean ball and $b=0$ the fibre is the intersection of the two centred ellipsoids $\nP{v}^2\le K^2$ and $\|Lv\|^2\le\varepsilon^2$. Lagrangian duality with multipliers $\lambda_1,\lambda_2\ge0$ gives
\[
V_\varepsilon(a)=\inf_{\lambda_1,\lambda_2}\Big[\lambda_1K^2+\lambda_2\varepsilon^2+\tfrac14\ipP{\rho_a}{(\lambda_1I+\lambda_2L^\ast L)^{-1}\rho_a}\Big],
\]
and minimisation over $\lambda_1$ at fixed $\tau=\lambda_2/\lambda_1$ gives the second form of \eqref{eq:Veps}. \qed

\paragraph{Proposition~\ref{prop:gram}} For fixed $\tau$, the quadratic $\nP{v}^2+\sum_j\tau_j\|L_jv-b_j\|^2$ has the minimum $c_\tau=b\T(\Gamma_{DD}+\Lambda_\tau)^{-1}b$ at $\hat v=L^\ast(\Gamma_{DD}+\Lambda_\tau)^{-1}b$ and the Hessian $2(I+L^\ast\Lambda_\tau^{-1}L)$. Every $v\in C$ satisfies $\nP{v}^2+\sum_j\tau_j\|L_jv-b_j\|^2\le K^2+\sum_j\tau_j\varepsilon_j^2$, so $C$ lies in an ellipsoid centred at $\hat v$, which is empty if the right-hand side is smaller than $c_\tau$. The maximum of $a\T DF\,v$ over this ellipsoid is $a\T m_\tau$ plus the square root in \eqref{eq:Vgram}, by the Woodbury identity $\mathcal D_F(I+L^\ast\Lambda_\tau^{-1}L)^{-1}\mathcal D_F^\ast=\Sigma_\tau$. This gives the upper bound for every $\tau$; equality of the infimum with the supremum, and of the feasibility condition with $K\ge K_{\min}$, is Lagrangian duality for the convex quadratic constraints, which holds under the interior-point assumption. \qed

\paragraph{Lemma~\ref{lem:activity}} The margin of constraint $t$ at $\psir+v$ is $m_t-\ipP{\gamma_t}{v}$, and the maximum of $\ipP{\gamma_t}{v}$ over the linearised fibre is $V_\E(\gamma_t)$ by definition. \qed

\paragraph{Theorem~\ref{thm:kkt}} Problem \eqref{eq:nlp} is a smooth programme with two inequality constraints in the Hilbert space $\X$. Under linear independence of the active constraint gradients the Karush--Kuhn--Tucker conditions hold \cite{bonnans2000}: $Df[\psi^\ast]=\alpha\,\ipP{\psi^\ast-\psir}{\cdot}+\beta\,(\Ob[\psi^\ast]-y)\T D\Ob[\psi^\ast]$ with $\alpha,\beta\ge0$. Passing to representers at $\psi^\ast$ gives \eqref{eq:kkt}. Condition \eqref{eq:sosc} states that the Hessian of the Lagrangian is negative definite on the tangent space of the active constraints, which is the second-order sufficient condition \cite{bonnans2000}. \qed

\paragraph{Theorem~\ref{thm:parity}} Write $\psi=\psir+tw$. With $\phi(w,t)=(f[\psir+tw]-f[\psir])/t$ and $\omega(w,t)=(\Ob[\psir+tw]-\Ob[\psir])/t$, extended to $t=0$ by the derivatives, the end points are $f^\pm(t)-f[\psir]=t\,\Phi^\pm(t)$, where $\Phi^+(t)$ and $\Phi^-(t)$ are the maximum and the minimum of $\phi(\cdot,t)$ over $S(t)=\{w:\nP{w}\le K_0,\ \|\omega(w,t)\|\le\varepsilon_0\}$. At $t=0$ this is the linear problem, whose maximiser $w_0$ is nondegenerate by assumption and whose minimiser is $-w_0$. By the implicit function theorem applied to the optimality system, the maximiser continues to a smooth branch $w^+(t)$ for $t$ in a neighbourhood of zero, of both signs, and $\Phi^+$ is smooth there. The functions satisfy $\phi(-w,-t)=-\phi(w,t)$ and $\omega(-w,-t)=-\omega(w,t)$, hence $S(-t)=-S(t)$ and $\Phi^+(-t)=-\Phi^-(t)$. With $\Phi^+(t)=\bar V+c_1t+c_2t^2+c_3t^3+\dots$ it follows that $f^+-f[\psir]=\bar Vt+c_1t^2+c_2t^3+\dots$ and $f^--f[\psir]=-\bar Vt+c_1t^2-c_2t^3+\dots$, which is \eqref{eq:parity} with $s_2=c_1$, $s_3=c_2$, $s_4=c_3$. \qed

\paragraph{Proposition~\ref{prop:inner}} The first statement holds because the supremum is taken over a subset. For a linear class, $v\in T$ implies $\ipP{\rho}{v}=\ipP{P_T\rho}{v}$, so the class problem is the linear problem with projected representers, and $\Gamma-\Gamma^\Theta$ is the Gram matrix of the vectors $(I-P_T)\rho_r$. For exact nominal data the class fibre is the ball of $T\cap\ker L$, so $V_0^\Theta=K\nP{P_{T\cap\ker L}\rho_a}$, and \eqref{eq:deficit} is the Pythagorean identity for $T\cap\ker L\subset\ker L$. The bound \eqref{eq:valueloss} is the second-order growth condition at a strict local maximiser \cite{bonnans2000}. \qed

\paragraph{Proposition~\ref{prop:outcome}} Let $C_0$ be the linearised fibre before the test and $H(a;y_c)=\sup\{\ipP{\rho_a}{v}:v\in C_0,\ |\ipP{\rho_c}{v}-y_c|\le\varepsilon_c\}$. The constraint set is jointly convex in $(v,y_c)$ and the objective is linear, so $H(a;\cdot)$ is concave, and so is $W(y_c)=\tfrac12(H(a;y_c)+H(-a;y_c))$. If $C_0=-C_0$, then $H(a;-y_c)=H(-a;y_c)$, hence $W(-y_c)=W(y_c)$, and a concave even function is largest at zero. \qed

\paragraph{Proposition~\ref{prop:limits}} (i) Since $H_\infty\subset\ker L$ for every design, $V\ge K\nP{P_{H_\infty}\rho_a}$. Conversely, for $\delta>0$ choose a finite combination with $\nP{P_{S_\infty}\rho_a-\sum_i\mu_i\rho_{\theta_i}}\le\delta$; once the design contains these observations, \eqref{eq:dual} gives $V\le K(\nP{P_{H_\infty}\rho_a}^2+\delta^2)^{1/2}+\max_i\varepsilon_i\sum_i|\mu_i|$, and the last term tends to zero with the tolerances. (ii) The perturbation $v=s\,\rho_a/\nP{\rho_a}$ with $s=\min(K,\varepsilon/M)$ lies in the ball and changes every observation by at most $sM\le\varepsilon$, so it is feasible and gives $\ipP{\rho_a}{v}=s\nP{\rho_a}$. (iii) If the sensitivity of every observation is a distribution of order at most $r$ supported in $\mathcal V$, it vanishes on every $v$ with $\partial^kv=0$ on $\mathcal V$ for $|k|\le r$; these $v$ form a closed subspace of $H_\infty$, and (i) applies. \qed

\paragraph{Propositions~\ref{prop:Arestr} to \ref{prop:Eprior}} Proposition~\ref{prop:Arestr} follows from the universal quantifier in the definition of validity and from the inclusion of images. For Proposition~\ref{prop:Bgain}, every post-test fibre is a subset of $\U$, which gives one direction; if $\psi\in\U\setminus\Rr$, the outcomes in $Y_c(\psi)$ retain $\psi$, which gives the other. For Proposition~\ref{prop:Cpair}, a common outcome of a consequential pair produces a fibre of mixed validity or of diameter larger than $\delta$, and conversely either defect of a post-test fibre exhibits such a pair. For Proposition~\ref{prop:Dcomplete}, if no $\U_N$ were contained in $\Rr$, the nested non-empty compact sets $\U_n\setminus\Rr$ would have a common point in $\U_\infty\setminus\Rr$; Hausdorff convergence of the images follows from continuity of $F$ on the compact set $\U_N$ and from nesting. Proposition~\ref{prop:Eprior} is proved in the text following it. \qed

\paragraph{Lemma~\ref{lem:gate}} Fix $\psi\in\mathcal C$, write $p=p(\psi)$ and $G(h)=\B(p+h;\psi)$ for $\|h\|_A\le R$. The lower bound on the tangent gives $\langle G(h_1)-G(h_2),h_1-h_2\rangle\ge\kappa\|h_1-h_2\|_A^2$. On the sphere $\|h\|_A=R$, $\langle G(h),h\rangle\ge\kappa R^2-\|G(0)\|_{A^{-1}}R>0$, so $G$ has a zero in the open ball \cite[Theorem~6.3.4]{ortega1970}, and it is unique by strict monotonicity. All states of the ball lie in $\Omega_c$ and the tangent at the solution is invertible, so $\psi\in\Rr$. \qed
\section{Exact certification of the gate for the plate}\label{app:cert}

This appendix summarises how the inequality of Lemma~\ref{lem:gate} is proved on the declared cell of Section~\ref{sec:plate}. The construction has four parts: a finite-dimensional comparison, an exact outer description of the feasible set, a moment relaxation that proposes a certificate, and an acceptance in exact arithmetic that does not depend on the solver.

\paragraph{Reduction to a scalar comparison} The laws are coordinates $u\in\R^{680}$ of the degree-14 space, subject to $\|u\|\le K$ in the declared metric, to four linear conditions that fix the natural state, and to the 120 coupon inequalities. Sixteen amplitude coordinates $a=U\T(u-u_c)$, with $u_c$ the minimum-norm law, are retained explicitly; the remaining 664 coordinates enter through uniform bounds. The predictor of the state is a low-order polynomial in $a$, and the operator $A$ of the gate is the tangent at the equilibrium of $u_c$. The residual of the predictor splits into a retained part, consisting of cubic and quartic terms in $a$ whose $A^{-1}$-norm is evaluated through the Gram matrix of $816$ cubic and $3876$ quartic finite-element image vectors, and remainder terms that are bounded uniformly on the cell. With the capacity $\kappa R$ evaluated for the cell, the gate on a parent box $\{a_0+p\circ z:|z_i|\le1\}$ takes the form
\begin{equation}\label{eq:closure}
\sup_{z\in\mathcal F\cap[-1,1]^{16}}Q(z)<K_0,
\end{equation}
where $\mathcal F$ is the image of the data-consistent set in the parent coordinates, $K_0$ collects the capacity less all constant charges, and $Q$ is an explicit polynomial of degree six, obtained from the retained residual by elementary inequalities that remove the square root. When the predictor is shifted to the parent, the comparison contains norms of energies, each of which is introduced as an auxiliary variable $s_j$ with the exact relation $s_j^2=g_j(z)$, and $Q$ becomes a polynomial in $(z,s)$.

\paragraph{Outer description of the feasible set} The set $\mathcal F$ is the projection of a convex set and is described from outside by supports, each valid by a duality argument and replayed in exact arithmetic from its multipliers: half-spaces from dual certificates of the convex programme, and balls $\|a+c_\mu\|\le r_\mu$ obtained from the norm constraint combined with non-negative multiples of the data inequalities. Supports are added where the comparison is largest: points of large $Q$ in the current outer set are projected onto $\mathcal F$ by a convex programme, and the separating half-space and ball are added if their exact replay excludes the point. Every support contains $\mathcal F$, so the outer set is an enclosure at every stage. The certificates use between about 460 and 1500 localising constraints per piece.

\paragraph{Moment relaxation} The supremum in \eqref{eq:closure} over the outer set is bounded by a moment relaxation of order three \cite{lasserre2001}. For the shifted comparison only the monomials in the auxiliary variables that $Q$ contains are lifted, which gives a moment matrix of size $1040$ with $98\,640$ moments in place of $1691$ and $195\,862$ for a complete lift. The relaxation is solved by a first-order conic method \cite{odonoghue2016}. Its output is used only as a proposal.

\paragraph{Exact acceptance} From the dual solution, floating-point factors $L_b$ and multipliers $\kappa_k$ are frozen, and the polynomial
\begin{equation}\label{eq:Xid}
X=\hat c+\sum_b\hat g_b\,\big\|L_b\T m_b\big\|^2+\sum_k\kappa_k\hat e_k
\end{equation}
is formed, where $\hat c=Q/\sigma$ with a power-of-two scale $\sigma$, $\hat g_b\ge g_b$ are floating-point polynomials dominating the exactly replayed supports on the box, $m_b$ are monomial vectors and $\hat e_k$ are the equality polynomials $s_j^2-g_j$ with their multipliers. On the feasible set the middle sum is non-negative and the equalities vanish up to their rounding, so $\hat c\le X+\sum_k|\kappa_k||\hat e_k-e_k|$. The polynomial $X$ is nearly constant by construction; it is bounded on the box by exact ranges of its monomials. The rounding error of forming \eqref{eq:Xid} is bounded a priori by the standard model of floating-point arithmetic \cite{higham2002} and summed in rational arithmetic. A piece is accepted if and only if the resulting bound is smaller than $K_0$ in exact arithmetic. The status and the objective value reported by the solver never enter, in the spirit of rigorous sum-of-squares certificates \cite{peyrl2008,rump2010}. Since only dual feasibility of the proposal matters, the solver is run with a scaling that favours it; with adaptive scaling the unresolved dual residual consumed five to six per cent of the capacity, with fixed scaling less than $0.3\%$.

\paragraph{Partitions and covers} A parent that does not close as a whole is partitioned along one or two coordinates or along an auxiliary variable, and the capacity is recomputed for each sub-box, which increases it. A cover check verifies in rational arithmetic that the pieces chain without gaps over the parent, that every interval endpoint used is proved, that each piece uses its own capacity consistently and that every margin is positive; it rejects deliberately corrupted covers, including gaps of $10^{-100}$. The eight parents are the leaves of a binary partition of the cell with seven exact splits, which is replayed by a separate integer-arithmetic implementation.

\paragraph{Scope} The acceptance, the supports and the cover are exact. The inputs, namely the cover, the transported law domain with the proved reference radius, the Gram matrix of the residual images, the capacities and the remainder bounds, are bound by cryptographic hashes but come from one computational chain; their independent reproduction remains to be done.

\end{document}